\documentclass[preprint,showpacs,amsmath,amssymb]{revtex4}
\pdfoutput=1
\draft
\usepackage{graphicx,color}

\usepackage{multirow}
\begin{document}
\title{
Density-functional investigation of rhombohedral stacks of graphene:
topological surface states, nonlinear dielectric response, and bulk limit}

\author{Ruijuan Xiao$^{1,3}$, F. Tasn\'adi$^{2}$, K. Koepernik$^{1}$,
J.W.F. Venderbos$^{1}$, M. Richter$^{1}$, and M. Taut$^{1}$} 

\affiliation{
{\small $^{1}$IFW Dresden, P.O. Box 270116, D-01171 Dresden, Germany}\\
{\small $^{2}$IFM,
Link\"oping University, S-581 83 Link\"oping, Sweden}\\
{\small $^{3}$Beijing National Laboratory for Condensed Matter Physics,
Institute of Physics, Chinese Academy of Sciences, Beijing 100190, China}
}

\date{{\small \today}}

\begin{abstract}
A comprehensive density-functional theory (DFT)-based investigation of
rhombohedral (ABC)-type graphene stacks with finite and infinite
layer numbers and zero or finite static electric fields applied
perpendicular to the surface is presented.
Electronic band structures and field-induced charge densities
are critically compared with related
literature data including tight-binding and DFT approaches
as well as with own results on (AB) stacks.
It is found, that the undoped
AB-bilayer has a tiny Fermi line consisting
of one electron pocket around the K-point and one hole pocket
on the line K-$\Gamma$.
In contrast to (AB) stacks, the breaking of translational symmetry
by the surface of finite (ABC) stacks produces a gap in the
bulk-like states for slabs up to a yet unknown critical thickness 
$N^{\rm semimet} \gg 10$,
while ideal (ABC) bulk ($\beta$-graphite) is a semi-metal.
Unlike in (AB) stacks, the ground state of (ABC) stacks 
is shown to be topologically non-trivial
in the absence of external electric field.
Consequently, surface states crossing the Fermi level must unavoidably 
exist in the case of (ABC)-type stacking,
which is not the case in (AB)-type stacks.
These surface states in conjunction with the mentioned
gap in the bulk-like states have two major implications.
First, electronic transport parallel to the slab is confined to
a surface region up to the critical layer number $N^{\rm semimet}$.
Related implications are expected for stacking domain walls and
grain boundaries.
Second, the electronic properties of (ABC) stacks are highly
tunable by an external electric field.
In particular, the dielectric response is found to be strongly nonlinear
and can e.g. be used to discriminate slabs with different layer numbers.
Thus, (ABC) stacks rather than (AB) stacks with more than two
layers should be of potential
interest for applications relying on the tunability by an
electric field.
\end{abstract}


\maketitle

\newpage

\section{Introduction}
Recently, much attention has been attracted by the class of
two-dimensional (2D) graphene systems,
which consist of one or a few layers of carbon 
atoms~\cite{novoselov04,neto09}.
Their unusual properties provided the impetus to
suggest many potential applications
as electronic materials~\cite{oostinga08,karpan07,nomura06}.
One of the key
properties of graphene devices is that their electric conduction
can conveniently be switched on and off. 
This aim has been achieved in several ways, 
including breaking the symmetry of a  graphene sheet by depositing it on a 
substrate~\cite{zhou07,giovannetti07}, controlling its band structure by 
doping~\cite{ohta06}, and adjusting the electronic properties of a
graphene bilayer by applying an
external electric field perpendicular to the 
surface~\cite{oostinga08,mccann06,castro07,min07}.
Among these methods, the latter
is most promising since the electric field can easily be 
manipulated and the process is reversible. It has been proven that 
the gap of a graphene bilayer can be tuned continuously from zero to 
mid-infrared energies by changing 
the strength of the field~\cite{castro07,min07}.

Besides the electric field, also the 
stacking sequence of graphene layers influences 
the electronic structure of this 
material \cite{latil06,aoki07,min08a,guinea06,koshino10}. 
Two stacking types, (AB) or Bernal stacking with hexagonal symmetry
and (ABC) stacking with rhombohedral symmetry,
exist besides a disordered fraction 
both in natural and in synthesized graphite~\cite{lipson42}.
Here and in the following we denote the general stacking
types with (AB) and (ABC), and specific stacks with AB (bilayer),
ABA or ABC (trilayer), etc.

Recent work using the tight-binding approach was
carried out to investigate the effect of the thickness of (AB)-type films
including single-layer graphene on their $\pi$-band overlap~\cite{partoens06}
and to study the effect of an external electric field on 
few-layer (AB)-graphene~\cite{lu06}.
Koshino~\cite{koshino10} compared the tight-binding
electronic structures of (AB) and (ABC) stacks in electric field
and found that a gap is opened in (ABC) stacks while (AB) stacks
with more than two layers stay semimetallic in the field.

Band structures of both (AB)- and (ABC)-type few-layer
systems obtained by density-functional theory (DFT) calculations
were examined by
Latil and Henrard~\cite{latil06} (without external electric field) and
by Aoki and Amawashi~\cite{aoki07} (with external electric field).
The electric-field response of the DFT band structure of an
ABC-trilayer was investigated by Zhang {\em et al.}~\cite{mcdonald10a}.
On the experimental side, the existence of
ABAB and ABCA stackings in tetralayer graphene was recently proved
by infrared absorption spectroscopy in combination with
tight-binding calculations; no other tetralayer stackings were
found~\cite{mak10}.

For bulk systems,
on the other hand, clear evidence that rhombohedral graphite is a true
carbon allotrope seems to be lacking, see Ref.~\onlinecite{cousins03} and
references therein. The available data point to the occurrence of
about ten-layers thick (ABC) stackings with areas in the order of
$10^5$ unit rhombi in filings from single crystals~\cite{cousins03}.
A comparative DFT study of (A), (AB) and (ABC) bulk systems was
performed by Charlier {\em et al.}~\cite{charlier94}.
Obviously, Bernal-type graphite ($\alpha$-graphite) is energetically 
more stable than rhombohedral graphite 
($\beta$-graphite)~\cite{charlier94,cousins03}
and has been studied much more intensely than the latter.

\begin{figure}
  \centering
  \includegraphics[width=0.8\textwidth]{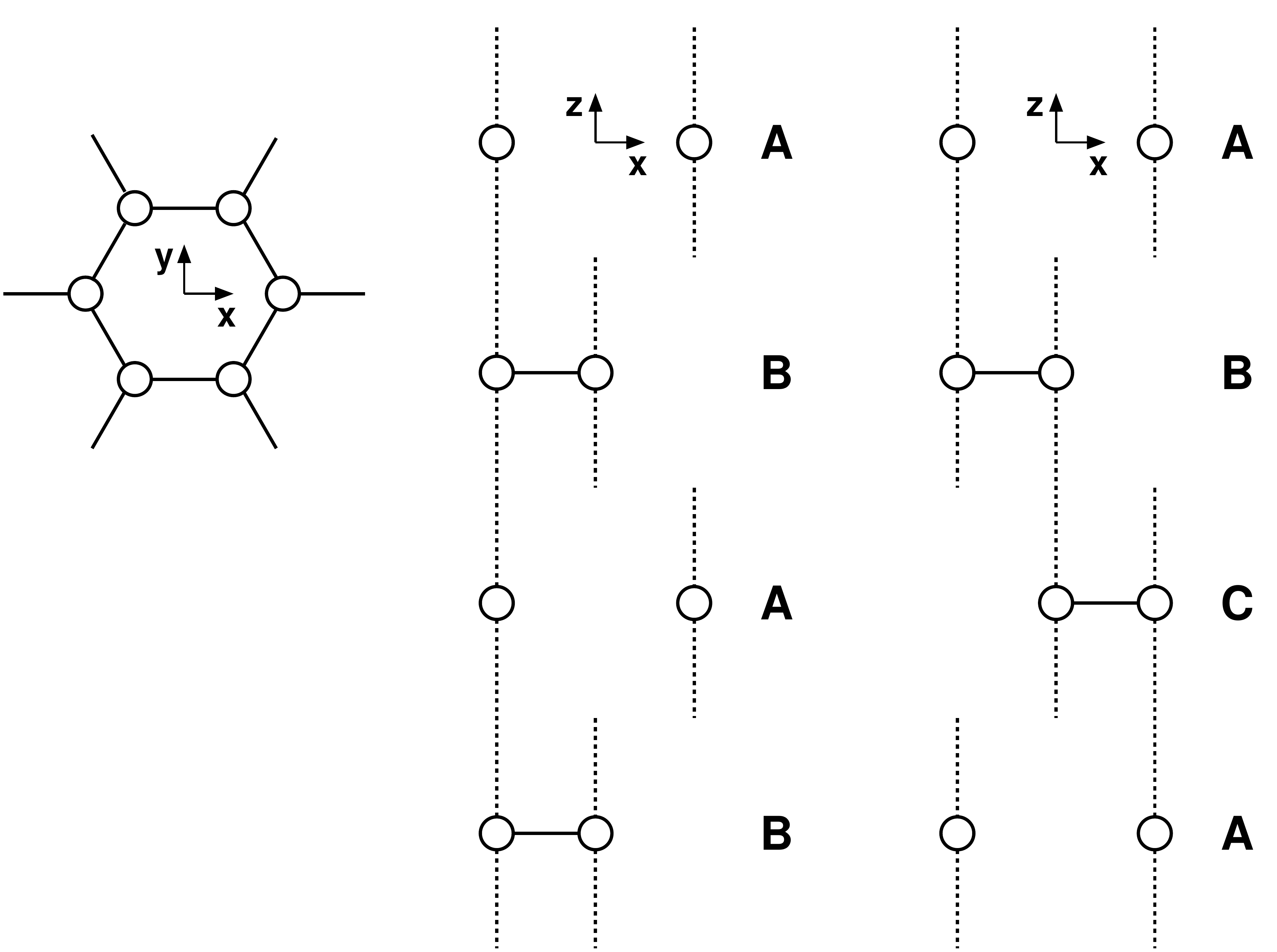}
  \caption{Sketch of the considered stackings of graphene.
Left panel: graphene layer; central panel: (AB) stacking;
Right panel: (ABC) stacking.
The (AB) stacking involves continuous lines of $p_z$ bonds for one 
kind of atoms and dangling bonds for the other atoms, while 
the (ABC) stacking is characterized by dimer-like $p_z$ bonds.}
  \label{fig:ab_vs_abc}
\end{figure}

The most important structural difference between (AB)- and (ABC)-type
stackings is that 
the (AB) stacking shows straight lines of $p_z$ bonds connecting 
all  atoms of one sub-lattice 
perpendicular to the layers through the whole stack, whereas 
in the (ABC) stacking only single pairs of atoms are connected  
perpendicular to the layers, see Fig.~\ref{fig:ab_vs_abc}.
In other words, in (AB) stacks there are two kinds of atoms,
one with dangling $p_z$ bonds in both $z$-directions and the other
without any dangling bonds,
whereas in (ABC) stacks each atom has one dangling $p_z$ bond, 
either in positive or in negative z-direction.
Evidently, this difference 
must have a large impact on the structure of the $\pi$-bands.

The present paper is devoted to a comprehensive study of the
electronic structure and the dielectric response of
stacks of graphene by means of high-precision
DFT calculations combined with a topological analysis at
tight-binding level.
While the focus lies on the hitherto less intensely studied
rhombohedral stacks including bulk $\beta$-graphite,
comparison with AB-bilayers and other Bernal stacks
is made in appropriate places below.

Computational details with emphasis on the correct treatment
of external electric fields applied to 2D slabs
are outlined in Section II.
Section III A contains a critical comparison of the AB-bilayer
band gap versus external electric field with related literature
data and the low-energy band structure of few-layer (ABC) systems
with and without field.
The low-energy states are characterized as surface states
in Section III B. Further, we provide evidence that
(ABC) slabs have a gap in their bulk-like states only up to
a critical thickness.
A detailed band structure of bulk
$\beta$-graphite is shown and discussed in critical
comparison with literature data.
In Section III C, we discuss the topological distinction between
rhombohedral stacks and Bernal stacks.
The former are shown to possess topologically
protected zero-energy surface states.
Next, the field-induced charge densities
of bi- and 6-layer systems and an analysis of induced
dipole moments for single- and bilayer systems are provided in Section III D.
The analysis of the electric field response is continued
in Section III E by comparing the field-induced charge transfer
with the chemical charge transfer, in Section III F where conclusions
for the modelling of the dielectric response of graphene stacks are drawn,
and in Section III G where local densities of states without and with
external field are shown.
The field-dependent dielectric constant of few-layer
stacks is presented in Section III H.
Finally, the main results are summarized and conclusions
are drawn in Section IV.
Earlier arguments by Aoki and Amawashi~\cite{aoki07}
concerning a better electric 
tunability of (ABC) stacks compared with (AB) stacks are confirmed
and extended.

\section{Computational details}

DFT calculations have proven to be powerful
in predicting the ground state properties of various systems,
including three-dimensional (3D)
bulk and 2D thin films without and with external electric field.
In this context, 2D films are frequently modeled by forming a 3D 
system from periodically stacked slabs with vacuum regions in between. 
There is however one severe problem with a straight
application of the supercell approach for 2D slabs in self-consistent
calculations with external fields.
The field-induced dipole moment of a single slab causes
a potential difference between the vacuum regions on
the two sides of the slab. This induced potential difference
would break the translational invariance of the supercell.
It is, thus, compensated by a linear potential term forced
by the periodic boundary conditions in a straight application.
This linear term is not harmful in the vacuum region but
it {\em reduces} the induced (screening) potential inside
the slab and thus {\em enhances} the effective external field
to a degree depending on the ratio between vacuum and slab thickness.

In order to avoid these problems, 
Kunc and Resta~\cite{resta83} applied a sawtooth-like external potential with 
symmetric teeth of alternately rising and descending flanks. 
This potential can be implemented by introducing 
alternately positive and negative capacitor plates into the vacuum regions.
In this model, 
two neighboring slabs form a double cell, 
where the dipole moments of both slabs cancel each 
other and the periodicity of the potential is saved.
To circumvent the extra numerical effort by doubling the supercell, 
Neugebauer and Scheffler~\cite{scheffler92} 
(see also a similar approach in Ref.~\onlinecite{bengtsson99}) 
proposed to use an asymmetric external sawtooth potential with 
ascending flanks followed by jumps. 
They introduced artificial dipole layers 
in the vacuum region to compensate the induced dipole moments.
Although  both methods produce a periodic potential,
which provides {\em in principle} the same electronic structure within 
the periodic slabs as within a single slab,
there are inaccuracies if the vacuum region is not wide enough, and 
the  extra  discontinuities in the effective  potentials
worsen the convergence of Fourier transformation-based 3D methods. 

Within the FPLO (full-potential local-orbital) approach~\cite{fplo},
Tasn\'adi recently implemented a single-slab method 
which guarantees the correct 
boundary behavior in the vacuum region~\cite{tasnadi07}. 
Like in similar isolated-slab methods~\cite{FLEUR},
the advantages of this approach are the following:
(1) there are no side effects of any artificial periodicity, 
(2) the external electric field is smooth everywhere, and 
(3) the electrostatic properties can be calculated correctly 
without any dipole correction.

Most of the presented results, in particular the data of
Table~\ref{table:excess-electrons} and of all figures except
Figs.~\ref{fig:BS-bilayer}, \ref{fig:dos-bilayer},
\ref{fig:3DBS-bilayer}, \ref{fig:bulk+10L},
\ref{fig:bulk-projected}, and \ref{fig:hexfromrtg}
were obtained with
the FPLO-slab-1.00-10 code~\cite{tasnadi07}.
The data of the other figures,
where no external electric field is considered, were calculated
with the FPLO-code~\cite{fplo,koepernik99}, version $9.01$,
in a 3D supercell approach.
The local density approximation (LDA)
with the exchange-correlation (XC) parameterization
by Perdew and Wang~\cite{Perdew92a} was applied in all calculations. 

Fixed geometries according to Fig.~\ref{fig:ab_vs_abc}
with $1.42$ \AA{} C-C bond length and $3.33$ \AA~\cite{note2}
inter--layer distance were used both in the 2D and 3D approaches.
Thus, possible surface relaxations and bucklings were disregarded.
On the other hand and at variance with tight-binding model calculations,
surface effects
on the charge distribution and influence of the stacking and the
surface on the hopping parameters are taken into account in
a self-consistent manner.
Further, since the present investigation is focussed on the electronic
structure and not on structural or elastic properties, the fixed
inter--layer distance avoids any functional-dependence of the geometry
which might be relatively large for weak bonds.
In the supercell calculations, a well-converged 
vacuum thickness of 16 \AA{} was chosen.
The following space groups were taken: 166 for bulk $\beta$-graphite,
143 for 3D supercells, and 69 (layer group p3m1) for 2D slabs.

The basis set comprised
carbon 1$s$, 2$s$, 2$p$, 3$s$, 3$p$, and 3$d$ orbitals.
It was enhanced by including $4d$ orbitals for the 
calculation reported in Fig.~\ref{fig:hexfromrtg} in order to obtain a precise
empty-states band structure.

If not stated otherwise, we used a {\bf k}-mesh of 
$300 \times 300$ points in the full Brillouin zone
(FBZ) for the slab calculations and
of $90 \times 90 \times 1$ points for the supercell calculations
in a linear tetrahedron method with Bl\"ochl corrections.

In order to cross-check the FPLO results and to elucidate the role 
of the dipole correction in 3D  supercell calculations,
we compared AB-bilayer results of the 
pseudo-potential code QUANTUM-ESPRESSO-3.2.3 (QE) 
in supercell geometry~\cite{espresso,gianozzi09} (both with and without 
dipole correction, Fig.~\ref{fig:DipoleCorrection}) with the FPLO slab results.
The same atomic positions and LDA version were used in QE and FPLO calculations.
In QE, a supercell height of $c=16$ \AA{}
(implying a vacuum thickness of $12.67$ \AA) was used.
An external sawtooth-like potential was applied with
linear increase between $z=-7.2$ \AA{} and $z=7.2$ \AA{}
(slab centered at $z=0$) and linear decrease in the remaining region.
The cut-offs used in QE are 40 Ryd for the wave functions and 480 Ryd
for the charge density. We obtained converged band gaps with 
a {\bf k}-mesh of $\rm 60\times 60 \times 1$ points
and a cold smearing of 0.0001 Ryd (Monkhorst-Pack sampling). 

\section{ Results and discussion}

\subsection{Field-dependent band structure of AB, ABC, ABCA, and ABCAB stacks}

\begin{figure}
  \centering
  \includegraphics[width=0.8\textwidth]{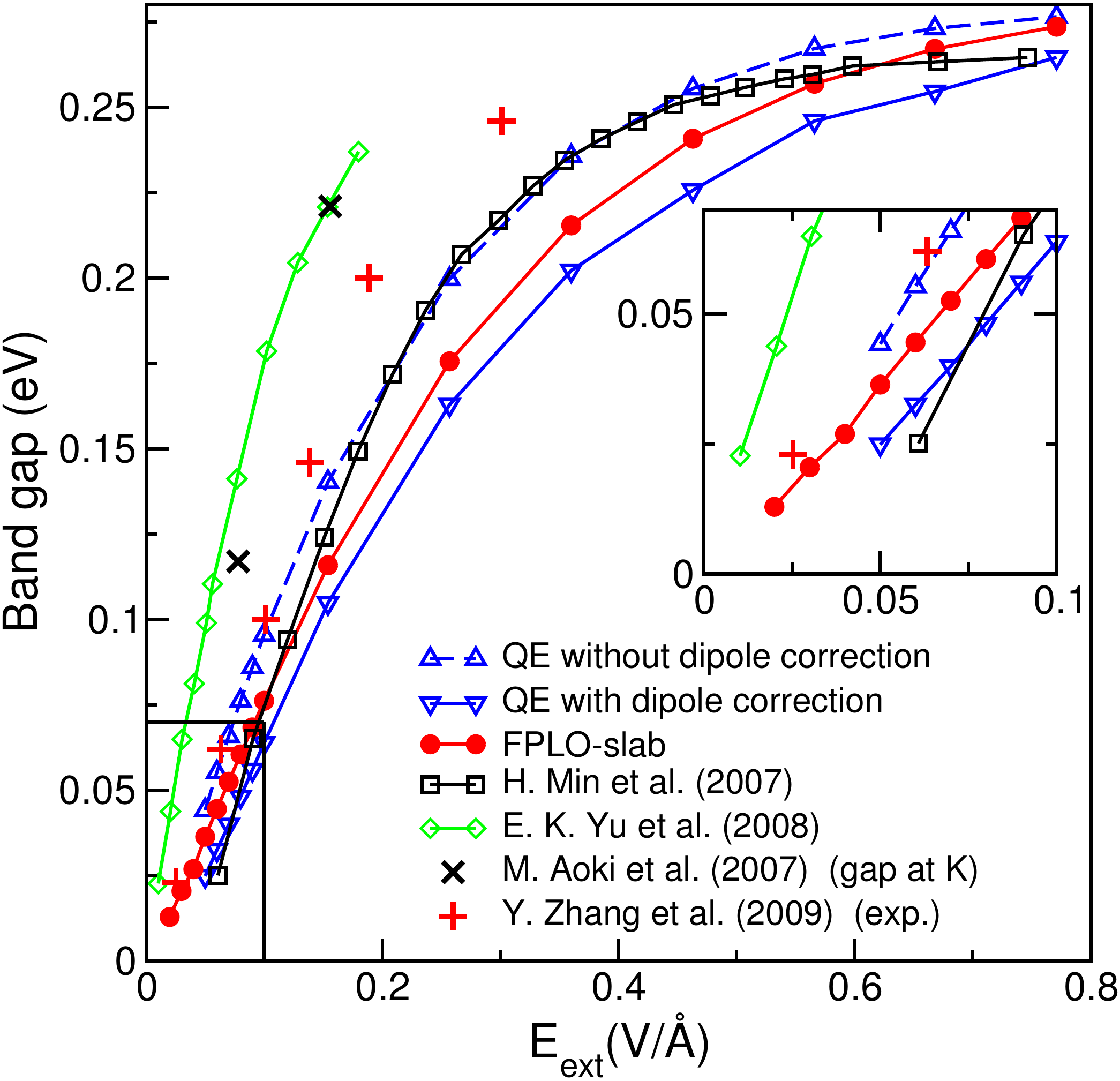}
  \caption{(Color online) Band gap of AB-bilayers versus external
field: Comparison of our FPLO--slab results
(bullets) with several supercell approaches and with 
experimental data (+ signs, Ref.~\onlinecite{zhang09}).
QE results (this work) are denoted by open triangles (up: without
dipole corrections, down: with dipole corrections), literature data
by squares (Ref.~\onlinecite{min07}), diamonds (Ref.~\onlinecite{yu08}),
and crosses (Ref.~\onlinecite{aoki07}), respectively.
Lines are intended to guide the eye.
The insert is a blow-up of the low-field region marked in the 
main figure.
All curves show the minimum gap except that by Aoki and Amawashi (2007), 
where the gap at the K-point is given.}
  \label{fig:DipoleCorrection}
\end{figure}
\begin{figure}
   \centering
   \includegraphics[width=0.9\textwidth]{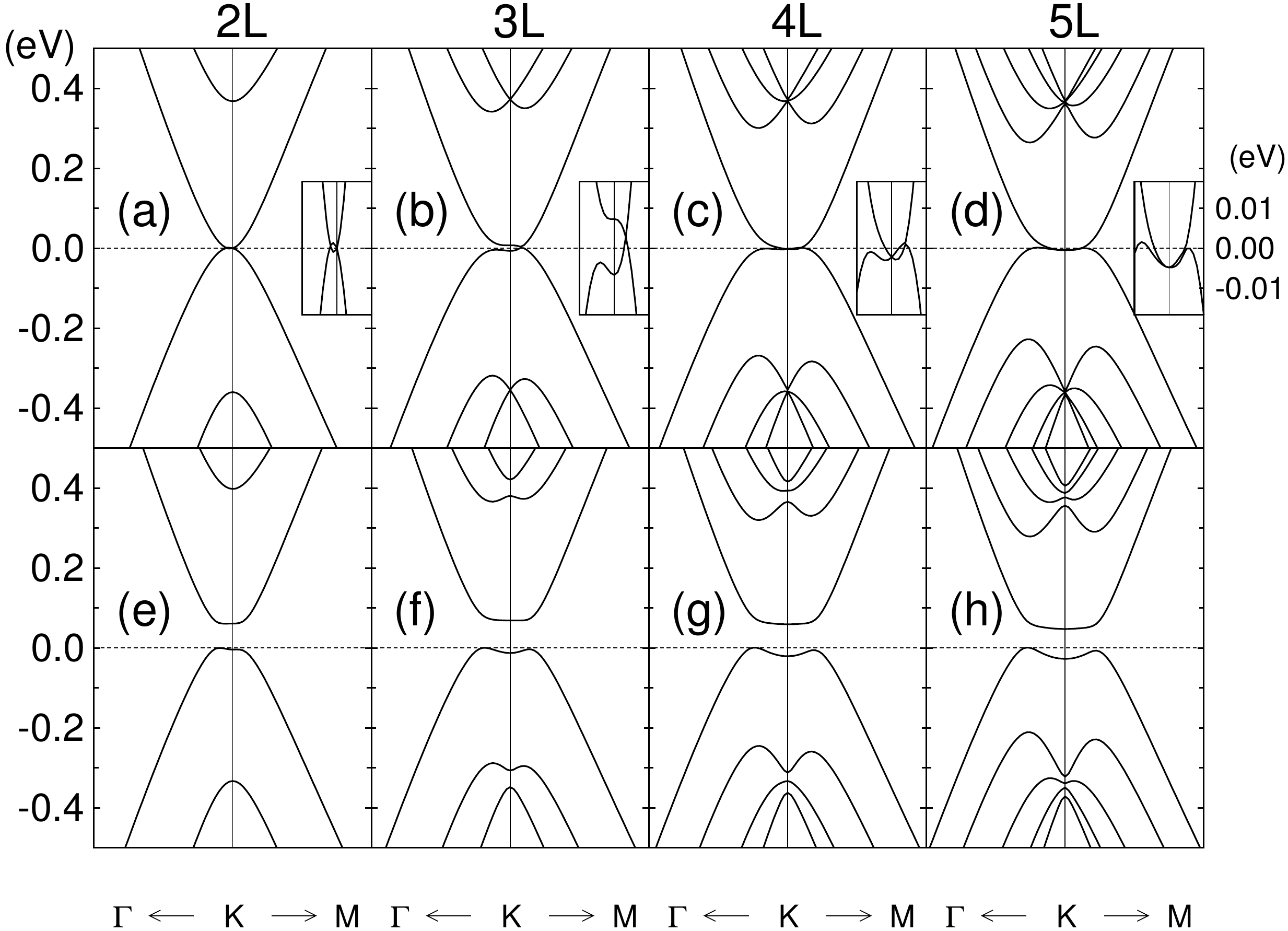}
   \caption{
DFT band structure for AB-bilayer (a, e), ABC-trilayer (b, f),
ABCA-tetralayer (c, g), and ABCAB-pentalayer (d, h).
Subfigures (a)-(d) refer to zero field and (e)-(h) to
$E_{\rm ext}=0.08$ V/\AA. 
The inserts show the subtleties around the Fermi level,
which lies at zero energy (dashed lines).
The fine but still limited {\bf k}-mesh yields a slightly
shifted position of the Fermi level in Subfigures (a) and (b).
A higher resolution for the bilayer case is given in the
following figures. In the case of the trilayer we note that
the Fermi level should be situated exactly at the band crossing.
The band structures are shown
close to the symmetry point K (vertical lines) along the lines 
K $\rightarrow$ M and K $\rightarrow \; \Gamma$ as indicated 
below the figure. In the main figures, 20\% of the line K-M and
10\% of the line K-$\Gamma$ are shown, while the inserts are restricted to
5\% of K-M and 2.5\% of K-$\Gamma$.
The energy scale on the left refers to the main
figures, the energy scale on the right refers to the inserts.
}
   \label{fig:BandStr}
\end{figure}
\begin{figure}
\includegraphics[width=0.75\textwidth,angle=-90]{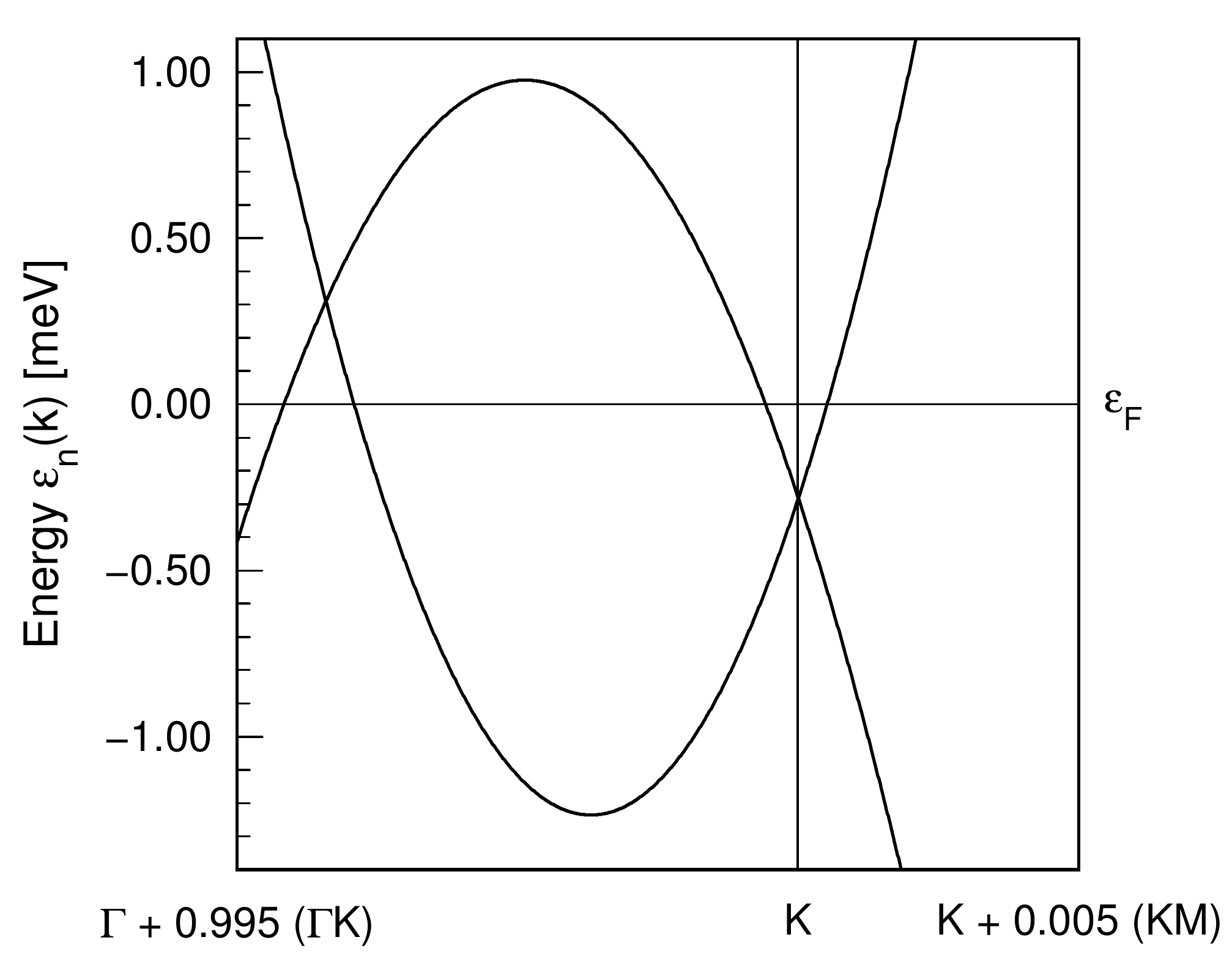}
   \caption{
Blow-up of the band structure of the AB-bilayer without external field
around the K-point.
The left end of the plot lies $0.5 \%$ of the distance K-$\Gamma$ 
from the K point toward the $\Gamma$ point, and the right end 
$0.5 \%$ of the distance K-M toward the M point.
}
   \label{fig:BS-bilayer}
\end{figure}
\begin{figure}
\includegraphics[width=0.6\textwidth]{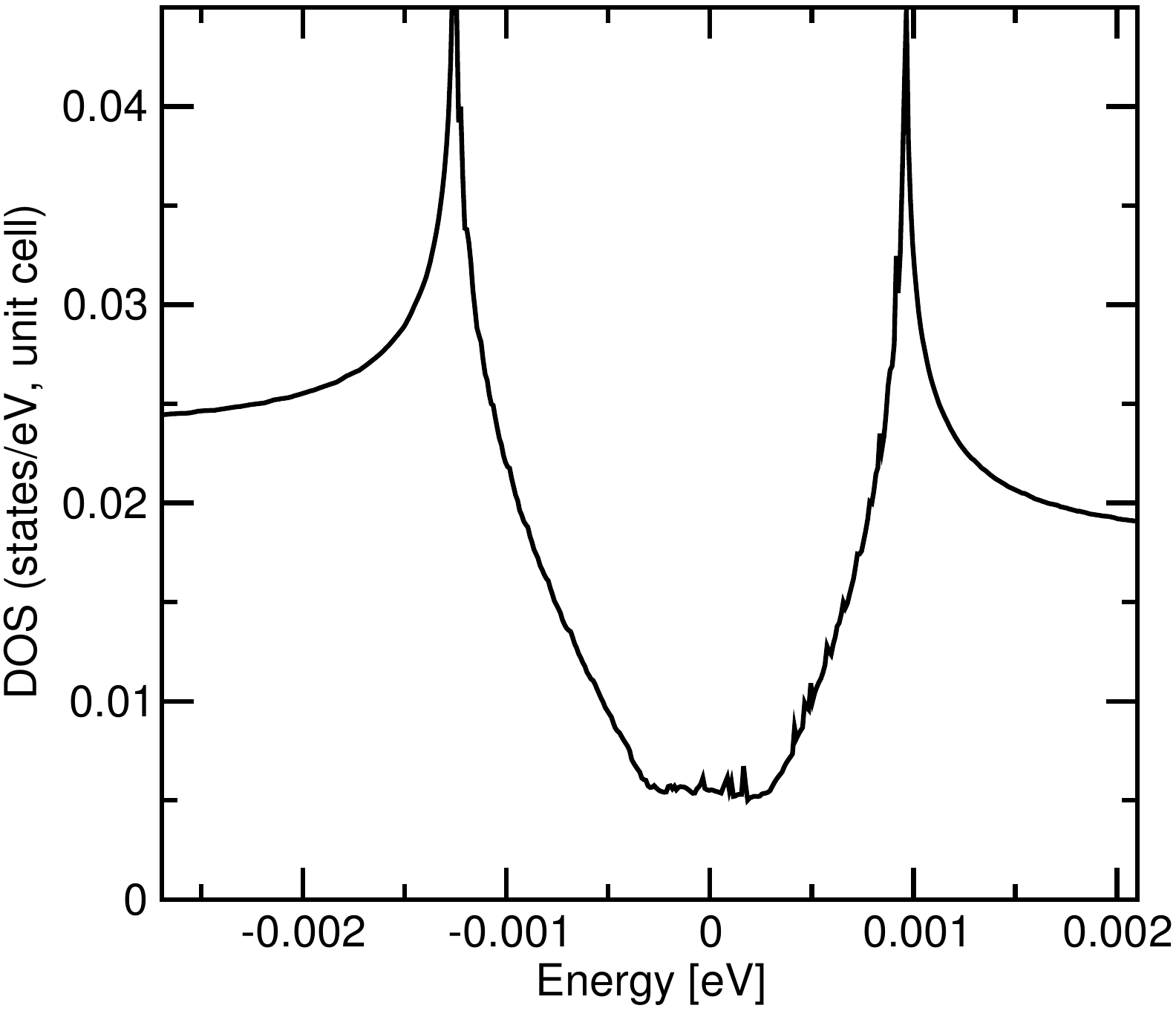}
   \caption{
Density of states of the AB-bilayer on the meV scale.
The minor peaks originate from the finite resolution of the {\bf k}-mesh.
}
   \label{fig:dos-bilayer}
\end{figure}
\begin{figure}
   \centering
   \includegraphics[width=0.9\textwidth]{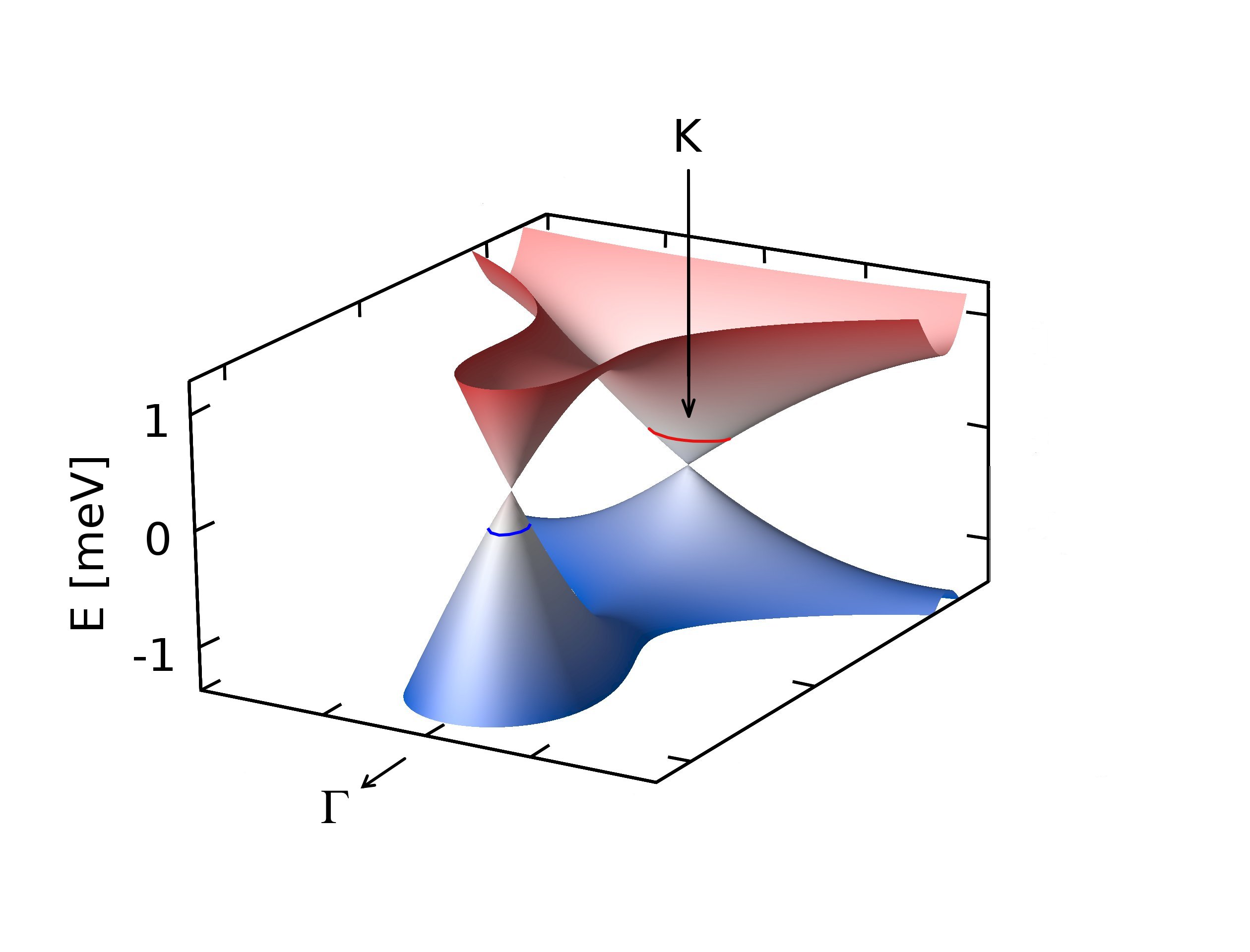}
   \caption{(Color online)
3D plot of the band structure of the AB-bilayer without external field 
around the K point. The Fermi lines of the  tiny electron (around K) and hole
(on the line K-$\Gamma$) pockets are indicated.
The data originate from the same calculation as those of 
Fig.~\ref{fig:BS-bilayer}.
Note, that the scaling of the two {\bf k}-axes differs by
a factor of about 2.
}
   \label{fig:3DBS-bilayer}
\end{figure}
\begin{figure}
  \centering
  \includegraphics[clip,width=0.7\textwidth]{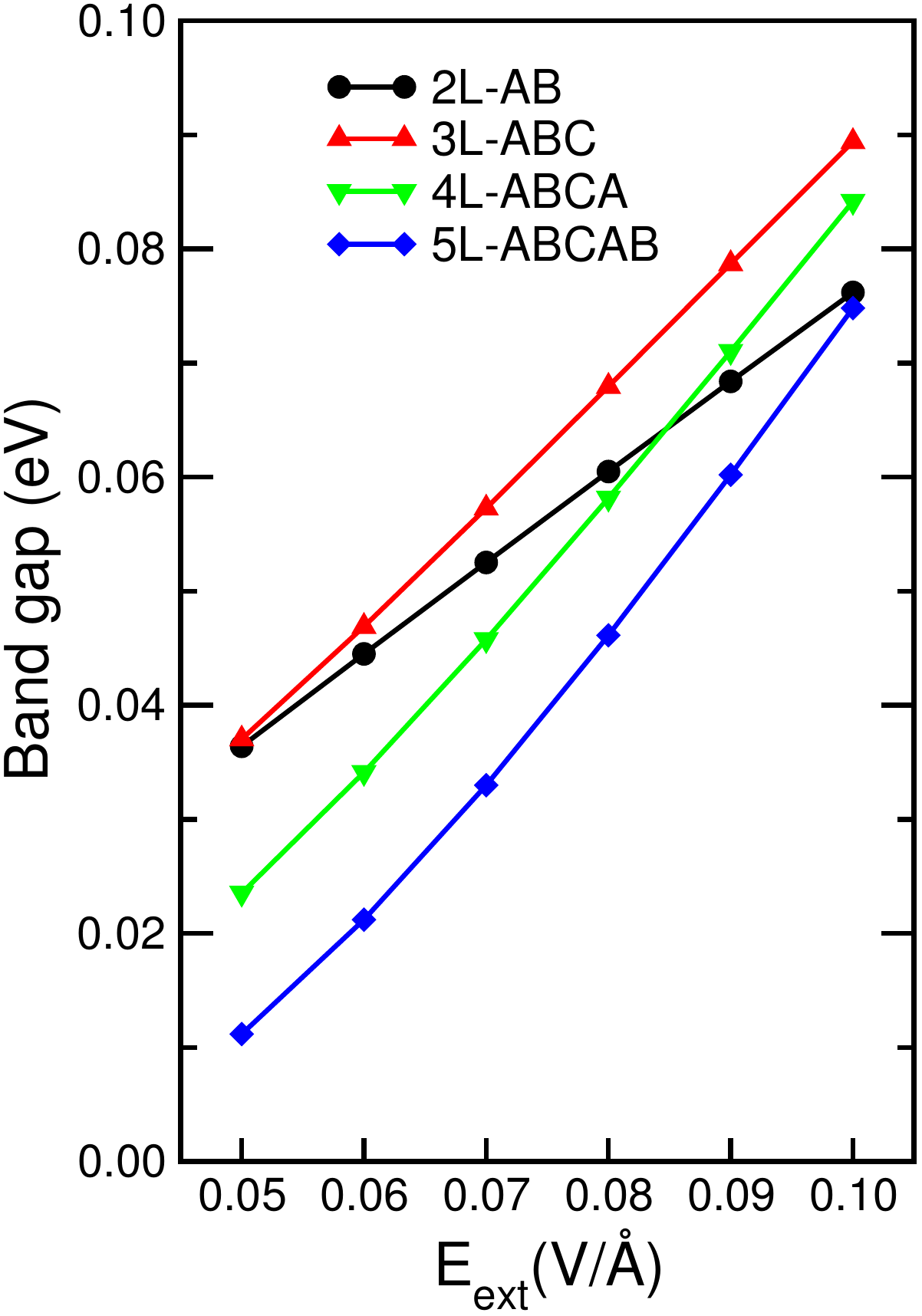}
  \caption{(Color online)
Minimum band gap for AB-bilayer and $3 \ldots 5$ layer
(ABC)-type graphene systems versus $E_{\rm ext}$.}
  \label{fig:GapvsE}
\end{figure}

Fig.~\ref{fig:DipoleCorrection} shows calculated band gaps 
of graphene AB-bilayers in an 
external electric field $E_{\rm ext}$ using different approaches and codes.
We first note, that
the dipole correction in QE
decreases the band gap substantially. 
According to our discussion in the previous section,
the induced screening field inside the slab
is artificially reduced by approximately a factor
(vacuum thickness divided by supercell height),
if the dipole correction is omitted,
thus enhancing the effect of the external field.
For the given choice of QE parameters and $E_{\rm ext}=0.1$ V/\AA, 
the band gaps obtained using QE with and without dipole correction 
amount to 64 meV and 96 meV, respectively.
The band gaps calculated with the FPLO-slab code, which does not need a 
dipole correction, are about 10$\ldots$15 meV larger than those obtained
with the QE-code with dipole correction.
All FPLO data fall in between the QE data with and without correction,
thus providing confidence in the new slab approach.
Notably, our and most of the other values are smaller than 
the experimental values deduced from infrared absorption~\cite{zhang09}.
This
can be attributed to the fact that DFT (even if it could be done exactly) 
systematically underestimates the Kohn-Sham gap width~\cite{hybertsen86}.

The energy distance between the two Dirac points
of the semi-metallic AB-bilayer is very small,
0.8 meV~\cite{latil06} or even less (see our results below).
Thus, a small but finite external electric field is needed to
open a gap in this system. Indeed, an extrapolation of the experimental
data in Fig.~\ref{fig:DipoleCorrection} yields a field of about
0.003 V/\AA{} necessary to open a gap.
This value coincides with the extrapolated FPLO-slab result.
On the other hand, extrapolation of the data by Min {\em et al.}~\cite{min07}
provides a critical field about one order of magnitude larger,
while the data by Yu {\em et al.}~\cite{yu08} point to a small gap
even at zero field (see insert in Fig.~\ref{fig:DipoleCorrection}).
In order to suggest possible reasons for
the severe deviations between our results and the
other published DFT data, we note the following problems.
In the calculation by Min {\em et al.}~\cite{min07},
a single bilayer was placed in a supercell without applying
a dipole correction. This could explain the similarity between
their data and the uncorrected QE data for intermediate fields.
Further, these authors used a relatively fine {\bf k}-mesh of
5,000 points in the 3D FBZ, which however might
still be too coarse at lower fields.
Note, that our calculations were carried out
with an ultra-fine mesh of 90,000 points in the 2D (sic) FBZ.
Yu {\em et al.}~\cite{yu08} employed a
perturbative bulk approach based on the modern
Berry-phase technique to investigate the effect of a homogeneous
electric field. To the best of our knowledge, this approach 
is also lacking the dipole correction for the study of slabs.
Moreover, this method yields for the (semi-)metallic case of bulk
$\alpha$-graphite an unphysical finite bulk polarization~\cite{yu08} 
which might point to problems of the method with small- and
zero-gap systems. These authors
applied a $k$-sampling with 900 points in the 2D FBZ.
Finally, also Aoki and Amawashi~\cite{aoki07} do not mention
dipole corrections and used a {\bf k}-mesh with 100 points in the
2D FBZ.

Fig.~\ref{fig:BandStr} shows band structures near the K point
for 2$\ldots$5 layers of  graphene 
with stackings AB, ABC, ABCA, ABCAB.
The first and the second row display the electronic structure without and with
external electric field, respectively.
The inserts in the upper row show some subtleties at the
meV scale with ten times magnified energy scale.
Corresponding figures for the (AB) stacked systems can be found in 
Ref.~\onlinecite{lu06} (tight-binding approach with
overlap integrals from bulk graphite) and in
Ref.~\onlinecite{latil06} (self-consistent DFT-calculations
with the ABINIT code).

Apart from details on the meV scale, visible in the magnifications
in Fig.~\ref{fig:BandStr}, the results from a 
self-consistently screened  minimum-basis tight-binding Hamiltonian 
taking into account only the interaction parameters
$\gamma_0$ and $\gamma_1$ from bulk graphite  
(see Fig.~4 in Ref.~\onlinecite{koshino10})
agree {\em qualitatively} with Fig.~\ref{fig:BandStr}.
Such a qualitative agreement holds even in the meV-range,
comparing our self-consistent DFT results for zero electric field
(inserts of Fig.~\ref{fig:BandStr}) with Fig.~5 of Ref.~\onlinecite{koshino09},
where the low-energy band-structure of rhombohedral stacks was obtained
in a tight-binding approximation taking into account the interaction
parameters $\gamma_0 \ldots \gamma_4$.
On the other hand, we note large {\em quantitative differences}
between the DFT results and self-consistently screened tight-binding
data. For example, the AB-bilayer band gap, Fig.~\ref{fig:DipoleCorrection},
is a factor of about two smaller than the related band gap shown in Fig.~6 of 
Ref.~\onlinecite{koshino10}.
A possible reason for these differences will be discussed in Section III F.

Before we further discuss the results for stacks with 3, 4, and 5 layers,
we have a look at the bilayer results with yet higher resolution,
Figs.~\ref{fig:BS-bilayer}, \ref{fig:dos-bilayer}, and \ref{fig:3DBS-bilayer}.
The band structure in Fig.~\ref{fig:BS-bilayer}
agrees qualitatively with
the DFT result published in Fig.~4a of Ref.~\onlinecite{latil06} where
the ABINIT code was used. Our energy difference between the local extrema 
-- called $\varepsilon_{psg}$ in Ref.~\onlinecite{latil06} -- is 2.2 meV 
compared with 2.6 meV in Ref.~\onlinecite{latil06}.
Also, the energy difference between the Dirac points amounts to 0.6 meV
(Fig.~\ref{fig:BS-bilayer}) vs. 0.8 meV (Ref.~\onlinecite{latil06}).
An important difference is, however, that
in Ref.~\onlinecite{latil06} the intersection of both bands (the ``Dirac
point'') at the K-point coincides with the Fermi energy,
leaving only a hole pocket on the line $\Gamma$-K. 
This is not correct because the volume of the electron- and hole-pockets
must be equal in neutral carbon stacks. 
In order to get an accurate Fermi energy 
we chose the $\bf k$-grid in such a way 
that  at both intersection points of the two bands 
mesh points for the linear tetrahedron method are located.
Then the  linearly interpolated bands, which are used  
for the calculation of the density of states,
reproduce the real band structure around the band intersections
as good as possible.
To be specific, a mesh with $2850 \times 2850 $ points in the 2D FBZ
was used to obtain the accurate position of the Fermi level
in Fig.~\ref{fig:BS-bilayer}.
In order to achieve reasonable resolution in Fig.~\ref{fig:dos-bilayer},
a special mesh of $180 \times 180$ points was employed.
This mesh was chosen such that it fills only
about 0.1\% of the 2D FBZ around the K-point, the only region
where states in the energy window of interest are present.

The simplest two-parameter tight-binding model for the bilayer 
(considering only $\gamma_0$ and $\gamma_1$)
provides a semi-metal with  two touching parabolas 
at the K-point~\cite{mccann06b, guinea06},
leaving no Fermi surface at all.
If additionally $\gamma_3$ is taken into account,
the bands show the two ``Dirac cones'', 
but both have the same energy (namely the Fermi energy) and there 
is no Fermi surface  without doping~\cite{mccann06b}.
Our result, Fig.~\ref{fig:BS-bilayer},
means that the prediction of a Lifshitz transition to a gap-less 
state~\cite{aleiner10}, which rests on a band structure model including 
only these 3 overlap integrals, has to be revisited.
On the other hand, the numerical tight-binding approach
in Ref.~\onlinecite{lu06} which considered the 7 most important
overlap integrals from bulk graphite 
shows an energy difference in  the two cusps, but in the opposite order
compared with the self-consistent DFT approaches 
(i.e. the crossing at point K is higher in energy than the other one).
In other words, Ref.~\onlinecite{lu06} provides a hole pocket 
around K unlike the electron pocket in our self-consistent DFT approach.
Consequently we have to state, that the
meV-features and the topology of the Fermi surfaces are very delicate
and tight-binding results using overlap parameters from bulk graphite may
not agree with the results  from self-consistent DFT calculations.
A similar conclusion was drawn for the
ABC trilayer case in Ref.~\onlinecite{mcdonald10a}.

Fig.~\ref{fig:3DBS-bilayer} shows the band structure of an AB-bilayer
close to the Fermi level in a 3D representation.
The hole pocket on the $\Gamma$-K line 
is nearly elliptical (with the axis along $\Gamma$-K 
being 3 times longer than perpendicular to it) and the electron pocket around 
K is roughly a circle with a slight deformation 
maintaining a three-fold symmetry 
according to the symmetry of the band structure at the K point. 
An estimate using a  model with linear dispersion and circular
(electron pocket around K) and elliptical (hole pocket on $\Gamma$-K)
iso-energy curves 
provides for the number of holes (which equals the number of electrons)
in the pockets the tiny value of $4\cdot 10^{-7}$ per unit cell, 
or equivalently, $1\cdot 10^{-7}$ per atom.
The density of states (DOS) at the Fermi level 
obtained from the same model 
is for both holes and electrons 
(which need not to be equal) approximately $4\cdot 10^{-3}$ states per eV
per unit cell. This agrees well with the numerical value seen in 
Fig.~\ref{fig:dos-bilayer}.
The peaks in the DOS of Fig.~\ref{fig:dos-bilayer} are actually 
logarithmic singularities resulting
from the saddle points between the two Dirac points
(see Fig.~\ref{fig:3DBS-bilayer}). The two kinks at the bottom of the 
valley between the peaks are located at the Dirac point energies.

We now resume the discussion of stacks with more than two layers,
Fig.~\ref{fig:BandStr}.
Without external field, the ABC stack is a zero-gap semi-conductor,
while the ABCA and ABCAB stacks
show a small overlap between the valence and the
conduction band near the K point. 
This confirms previous DFT results on ABC stacks~\cite{latil06,mcdonald10a}
and on ABCA stacks~\cite{latil06}.
These findings together with the information that also 
(i) all other rhombohedral
stacks with up to ten layers and bulk $\beta$-graphite
(see Section III B) are found to be semi-metallic in our DFT calculations
and (ii) stacks with
$4 \ldots 7$ layers were found to be
semi-metallic in a tight-binding approach with
five hopping parameters~\cite{koshino09} leads us to claim that except the
trilayer {\em all} rhombohedral stackings are semi-metallic.

As a general feature, we notice in Fig.~\ref{fig:BandStr} that
for an arbitrary number of layers $N$,
there are only two bands near the Fermi level,
and $N-1$ bands around $\pm0.4$ eV. 
These two bands at the Fermi level were identified in earlier tight-binding
calculations as surface states~\cite{guinea06,min08a}.
When an external electric field is applied, the conduction
band becomes flat and the valence band becomes a little concave at the K point.
Further, the degeneracy around $\pm0.4$ eV is lifted.

Fig.~\ref{fig:GapvsE} shows the calculated minimum gap 
for the AB-bilayer and for $3\ldots 5$ graphene 
layers with (ABC)-type stacking as a function of the external electric field 
$E_{\rm ext}$ ranging from $0.05$ to $0.10$ V/\AA. 
Due to the subtleties in the band structure around the Fermi energy,
our self--consistent procedure did not converge 
for the 3$\ldots$5 layer systems in
small (but non--vanishing) external fields below approximately $0.05$ V/\AA.
Our results show that all considered systems 
have a band gap in the considered fields, 
though they are semi-metallic at zero field.
In all cases the band gap increases with $E_{\rm ext}$ 
in the considered range.

The prediction that a gap is opened for (ABC) stacks
by an external electric field
differs from the behavior reported for systems with
Bernal stacking and $N>2$.
There, diverse results were obtained by tight-binding calculations.
While Lu {\em et al.} found that small gaps for ABA- and ABAB-slabs
exist in finite external field ranges (see Fig.~4 of Ref.~\onlinecite{lu06}),
Koshino~\cite{koshino10} reported that these systems and also (AB) stacks
with more layers stay semimetallic in a moderate field.
DFT calculations for ABA- and ABAB-slabs support the latter 
result~\cite{aoki07}.

For the AB-bilayer the general trend in the gap width with external field
has been experimentally verified~\cite{zhang09}.
Also, field-induced band gaps were obtained for (ABC) tri- and tetralayers 
by means of earlier DFT calculations~\cite{aoki07,mcdonald10a}.

Extrapolation of the band-gap vs. field curves of Fig.~\ref{fig:GapvsE}
to zero field should yield a zero band gap for the trilayer and a very
small negative band gap of about $-1 \dots -5$ meV for the other systems,
corresponding to the band overlaps visible in Figs.~\ref{fig:BandStr} and
\ref{fig:BS-bilayer}.
While this can be achieved by an almost linear extrapolation for the
bilayer system, non-linear behavior is expected from our data for all
other systems. Thus, we predict a positive curvature of the low-field
band-gap vs. field dependence for rhombohedral stacks with three and more
layers, where the non-linearity increases with the number of layers.

\subsection{Bulk limit and surface states} 

\begin{figure}
  \centering
  \includegraphics[width=0.9\textwidth,angle=-90]{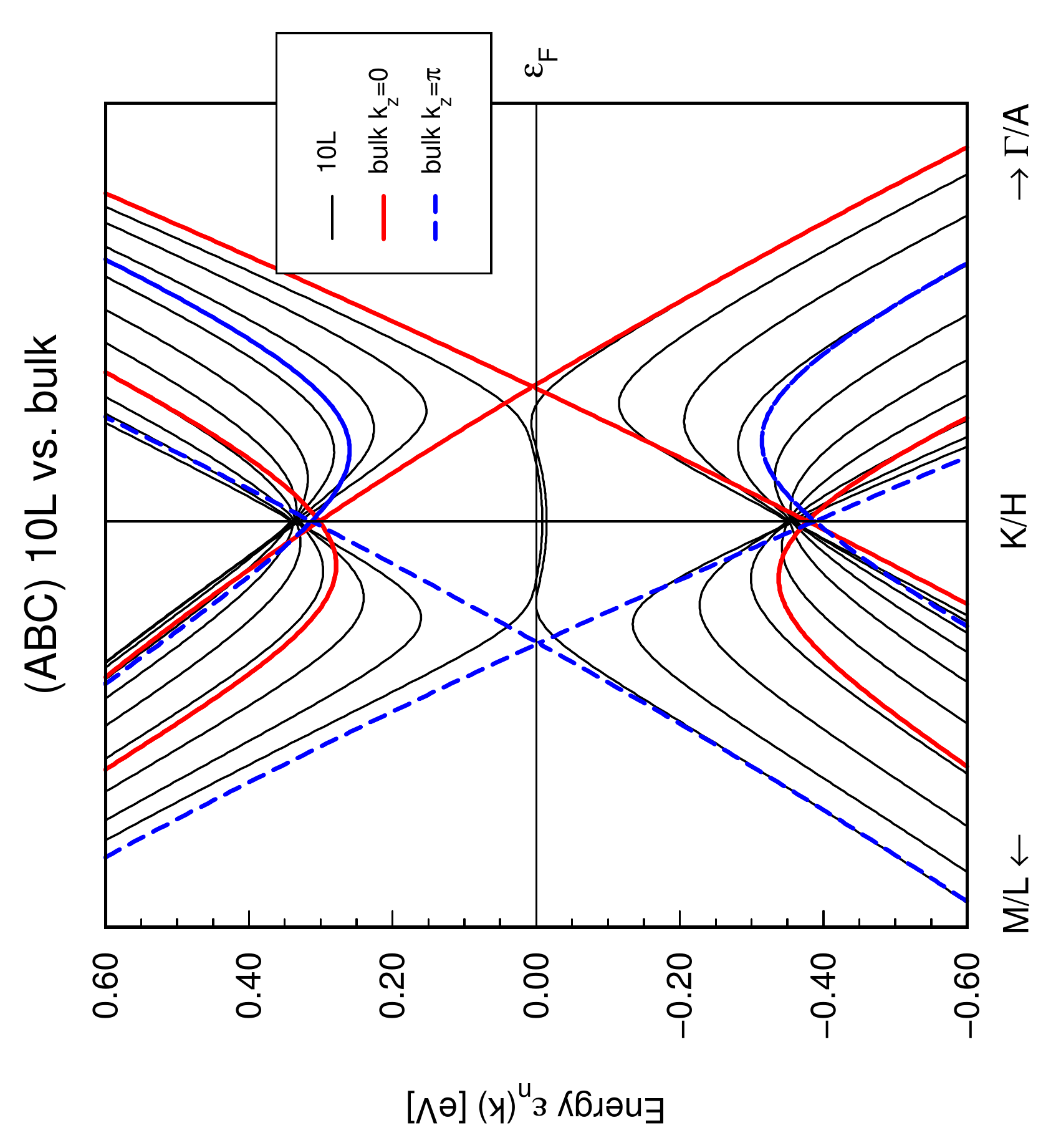}
  \caption{(Color online) Energy bands around the K-point of a
10-layer (ABC) stack (black, thin lines) together  
with the (ABC) bulk bands on a part of the lines M-K-$\Gamma$ (red, thick lines) 
and L-H-A (blue, dashed lines) 
lying on the bottom- and the top-plane 
of the hexagonal Brillouin zone.
Along K-M and H-L about 24\% of the whole line are shown and
along K-$\Gamma$ and H-A about 12\%.
The Fermi level of the bulk system is adjusted by 18 meV
to get the correct balance between electron and hole pockets.
For the sake of an easier comparison of slab- and bulk calculations,
the bulk calculation has been done in a hexagonal unit cell comprising 6 atoms,
although a smaller rhombohedral cell with 2 atoms exists.  }
  \label{fig:bulk+10L}
\end{figure}

\begin{figure}
  \centering
  \includegraphics[width=0.8\textwidth,angle=-90]{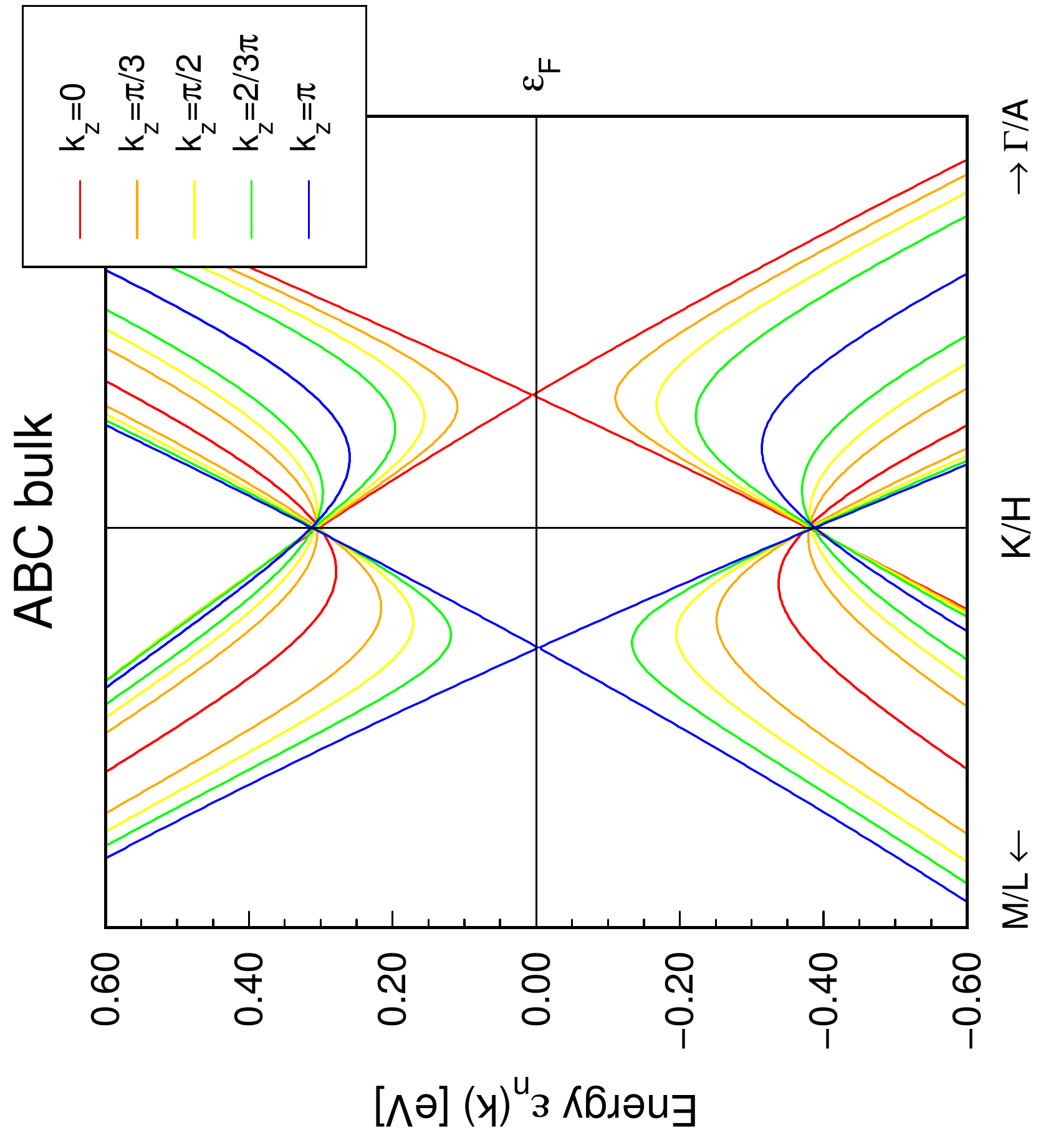}
  \caption{(ABC) bulk bands in the vicinity of the line K-H 
            projected onto 
            the M-K-$\Gamma$ plane in the Brillouin zone. 
            Five discrete $k_z$ values were adopted as shown in the legend. 
Along K-M and H-L about 24\% of the whole line are shown and
along K-$\Gamma$ and H-A about 12\%.
The Fermi level is adjusted by 18 meV to get the correct balance between
electron and hole pockets.
            The blue and the red curves (from the top- and the bottom-plane
            of the hexagonal Brillouin zone, respectively) 
            with the parabola-like maxima/minima are two-fold degenerate
            in addition to spin degeneracy.  
           }
  \label{fig:bulk-projected}
\end{figure}

An intriguing trait of the electronic structure of
finite (ABC) stacks is that 
all states in the vicinity of the Fermi energy are localized near the two 
surfaces. 
This was recently demonstrated for stacks of arbitrary thickness
by tight-binding model calculations~\cite{koshino09}.
Here we show by DFT calculations,
that there must be a finite thickness beyond which
the bulk gap is closed and the whole slab turns semi-metallic.

We first discuss the DFT results for slabs of finite thickness and for bulk
rhombohedral graphite.
In addition to the results on up to five layers,  Fig.~\ref{fig:BandStr},
consider the 10-layer result (for $E_{\rm ext}=0$)
shown in Fig.~\ref{fig:bulk+10L}. 
We observe in all stacks a pair of bands at the Fermi level
which is very flat over an increasing part 
of the {\bf k}-space around the point K, 
if the number of layers $N$ is  increased. 
The detailed behavior is complicated for thinner slabs with $N \le 5$
as seen in Fig.~\ref{fig:BandStr} and also for $N=6$, not shown. 
For $N\ge 7$ it however resembles qualitatively the behavior seen in 
Fig.~\ref{fig:bulk+10L} 
showing two  nearly parallel lines over a large {\bf k}-space region.
The Fermi energy lies amidst these two almost parallel bands. 
This behavior is a refinement of the $k^N$ dispersion provided
by the simplified model in Ref.~\onlinecite{min08a}.
Tight-binding calculations taking into account
the interaction parameters $\gamma_0 \ldots \gamma_4$ show
the same behavior for $N < 7$. 
Unlike our results, these calculations
still show crossings of the two bands for $N=7$,
Fig.~5 of Ref.~\onlinecite{koshino09}, but the 
band overlap is considerably smaller than for lower $N$.

A crucial point is that the wave functions belonging to the states 
of close proximity to the Fermi energy are strongly localized at the
slab surfaces, whereby each band is localized on one surface.
At the K symmetry point, 99\% of the related band weight belongs
to the top or surface layer.
If we follow these two bands away from the K point to the $\bf k$ 
regions where the splitting becomes increasingly larger,
they gradually turn into bulk-like states.  
The weights of all other bands,
which do not intersect the Fermi level, are spread all over the slab,
i.e., they are bulk-like.
This interpretation is supported by consideration of the 
projected bulk band structure (PBBS,
projected onto the plane parallel to the slab)
shown in Fig.~\ref{fig:bulk-projected}.
The PBBS rests on periodic boundary conditions in all 3 dimensions.
Concerning the relation between the PBBS 
and slab band structures for $N \rightarrow \infty$, 
the following theorems are familiar from surface band theory: \\
(i) The modulus of all slab states turns asymptotically 
(in inward direction) either into 
periodic bulk states or into surface states
with an exponentially decaying envelope.\\
(ii) The energy of the bulk states agrees exactly with the
corresponding states in the PBBS.
Therefore, all bulk bands lie within the continuum of the PBBS. \\
(iii) For given ${\bf k}_{||}$, the component of the Bloch vector parallel
to the slab, all states in gaps of the PBBS are surface states. 
Fig.~\ref{fig:bulk-projected} shows that the boundaries of the PBBS
in the central rhombic gap region are identical with the Dirac-like bands 
for $k_z = 0$ and $k_z = \pi$, also shown in Fig.~\ref{fig:bulk+10L}.
There, we can focus our attention to the ${\bf k}$ space around the line K-H,
because in all other regions the bulk bands are far away from the Fermi level. 
It follows from the above mentioned theorems, 
that the pair of states in the central rhombus of Fig.~\ref{fig:bulk+10L} 
have to be surface states, provided, this pair stays there in the limit 
$N \rightarrow \infty$. 
This conclusion coincides with the mentioned investigation of the 
weights of the Bloch states.
Our discussion also explains, why the pair of surface bands
turns into bulk bands upon immersion into the PBBS.

The bulk-like states of the 10-layer slab in Fig.~\ref{fig:bulk+10L} are 
not yet completely localized within the boundaries of the PBBS.
However, we checked that 
with increasing $N$ the energies of the bulk-like states at K
move closer to the Fermi energy making a complete 
overlap for $N \rightarrow \infty$ very likely.

It follows from Fig.~\ref{fig:bulk-projected},
that the low-energy band structure of bulk $\beta$-graphite
consists of four bands with linear (Dirac-like)
dispersion.
Two of these bands cross the Fermi level close to the
K-point, the other two close to the H-point.
This crossing is not lifted by spin-orbit interaction.
The ``Dirac points'' on the planes $k_z=0$ and $k_z=\pi$
have an energy difference of 9 meV and
lie above and below the Fermi energy, respectively,
giving rise to tiny hole and electron pockets
in the bulk band structure.
Thus, bulk $\beta$-graphite is a semi-metal. From the fact that
the Dirac points for $N \rightarrow \infty$
are found slightly below and slightly
above the Fermi level we conclude, that the gap of the
bulk-like states already closes at a large but finite stack
thickness, $N^{\rm semimetal} \gg 10$.

\begin{figure}
\noindent \begin{centering}
a)\includegraphics[clip,height=3.6cm]{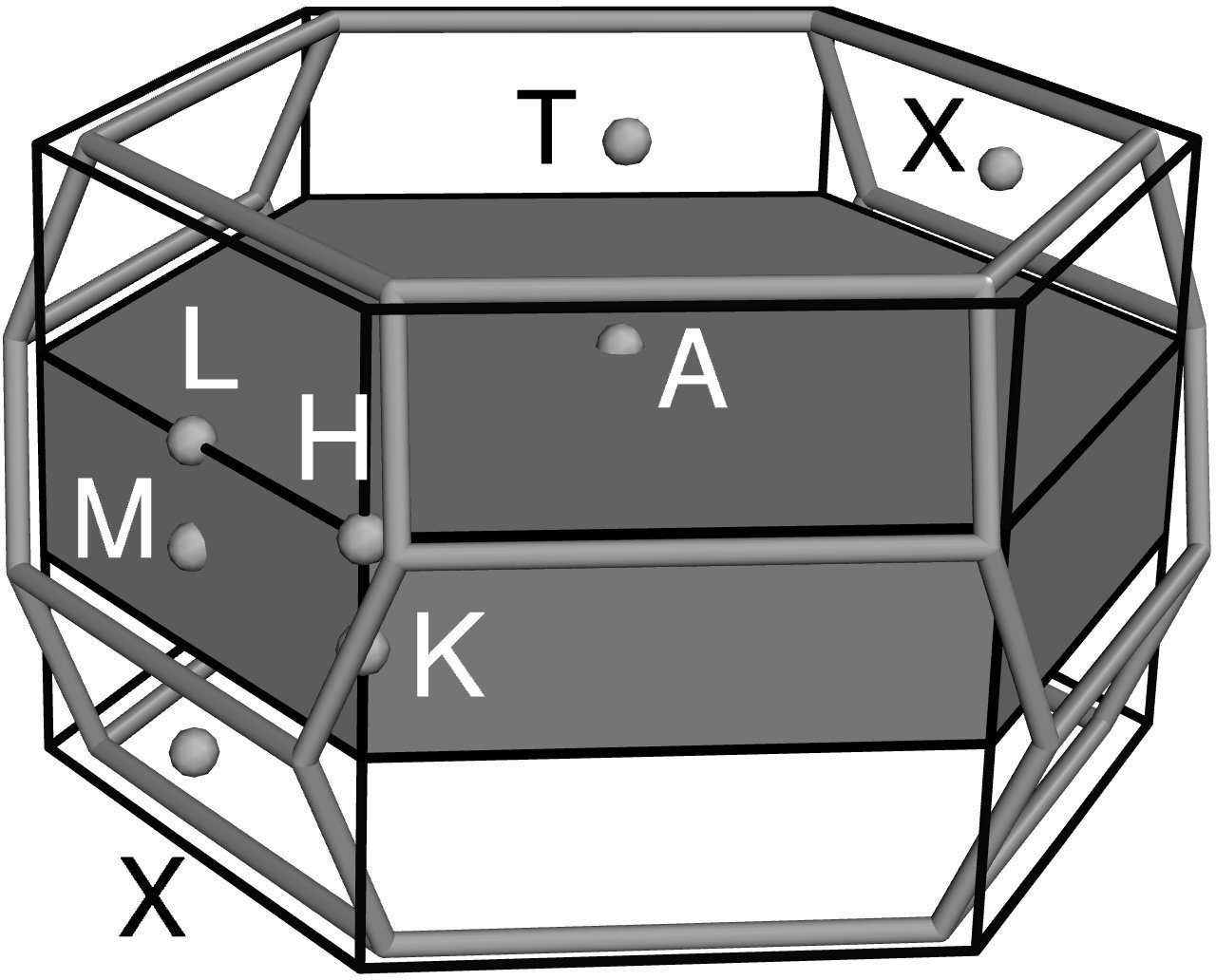}
\hspace*{5mm}
b)\includegraphics[clip,height=4.8cm]{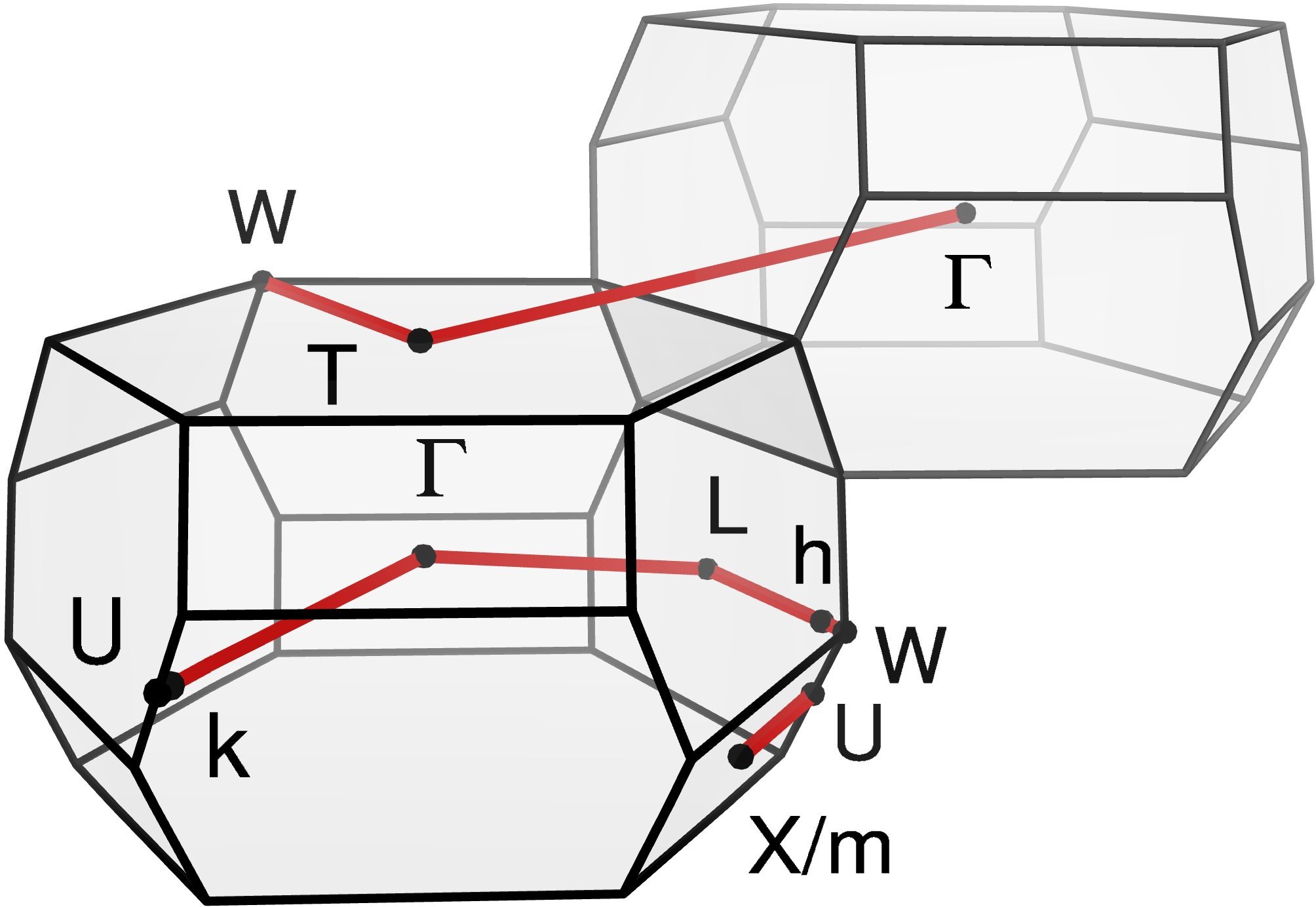}
\par\end{centering}

\caption{\label{fig:BZ-of-ABC}Brillouin zone
of (ABC) graphene. a) Three Brillouin zones
of the corresponding real-space conventional 6-atom
hexagonal cell are shown
together with the rhombohedral trigonal (RTG) BZ. High-symmetry
points of the hexagonal cell are denoted with white labels
and the points X and T of the RTG cell are denoted with black labels.
Note that the RTG and the hexagonal L-point
are identical. The RTG X-point falls onto the hexagonal M-point of
the translated hexagonal cells, which means that bands seen at X in
the rhombohedral band structure will be folded onto M in the hexagonal
setup, which contains three times as many atoms as the RTG setup.
In the same way the
RTG T-point is folded onto the hexagonal A-point.
b) The path through
the RTG BZ as used in the rhombohedral band structure of
Fig.~\ref{fig:hexfromrtg} (upper panel) is shown. Capital
labels denote rhombohedral high symmetry points and lower case
labels denote symmetry points in the hexagonal BZ.
Note that the paths X-U-k-$\Gamma$
and L-h-W-T form straight lines when continued into adjacent BZs.
The segment T-$\Gamma$ does not cut the BZ boundary in any high symmetry
point and was added to compare with the band structure
published in Ref.~\onlinecite{charlier94}.}

\end{figure}

\begin{figure}
\noindent \begin{centering}
\includegraphics[clip,angle=-90,origin=c,width=0.65\textwidth]{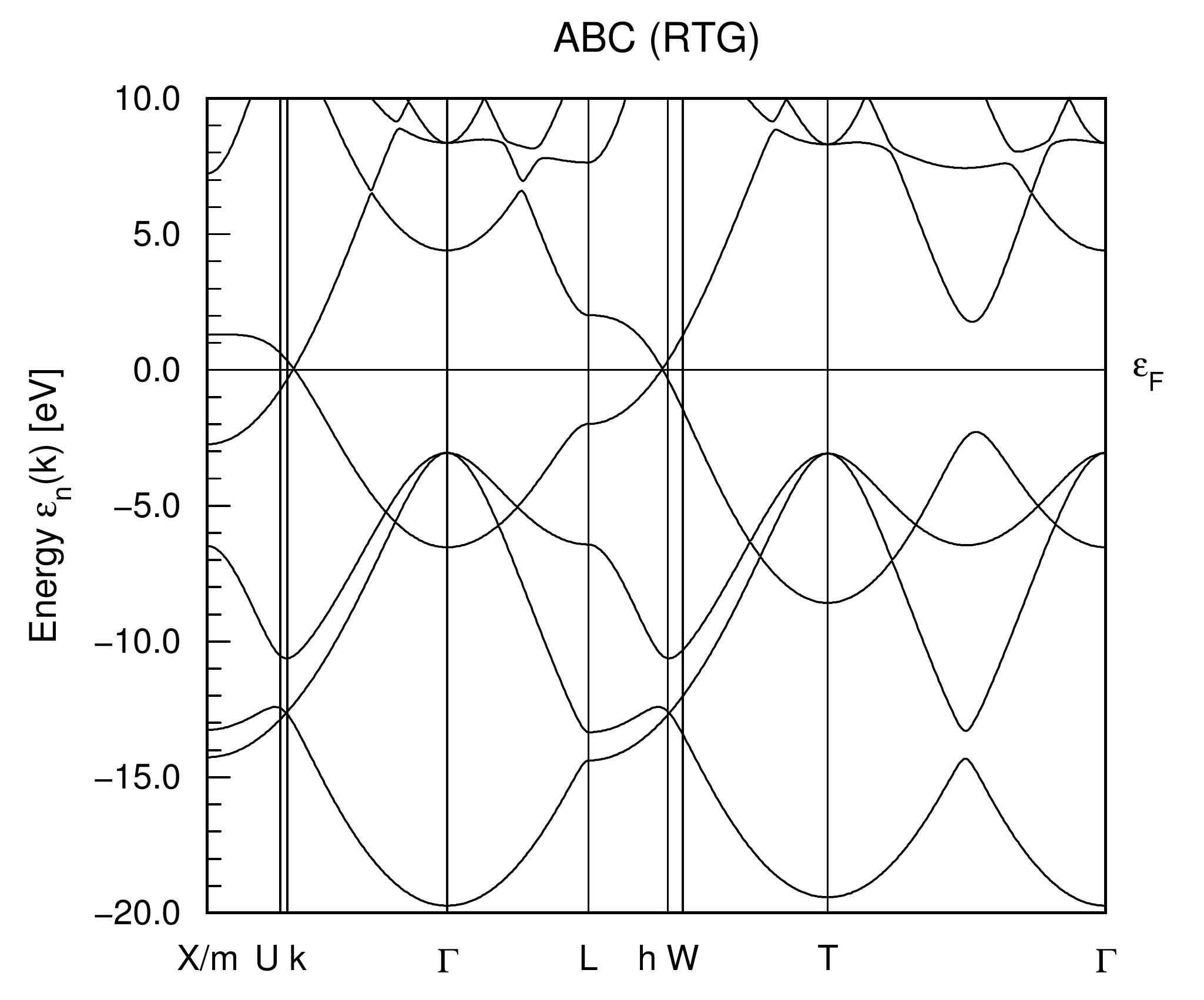}
\includegraphics[clip,angle=-90,origin=c,width=0.65\textwidth]{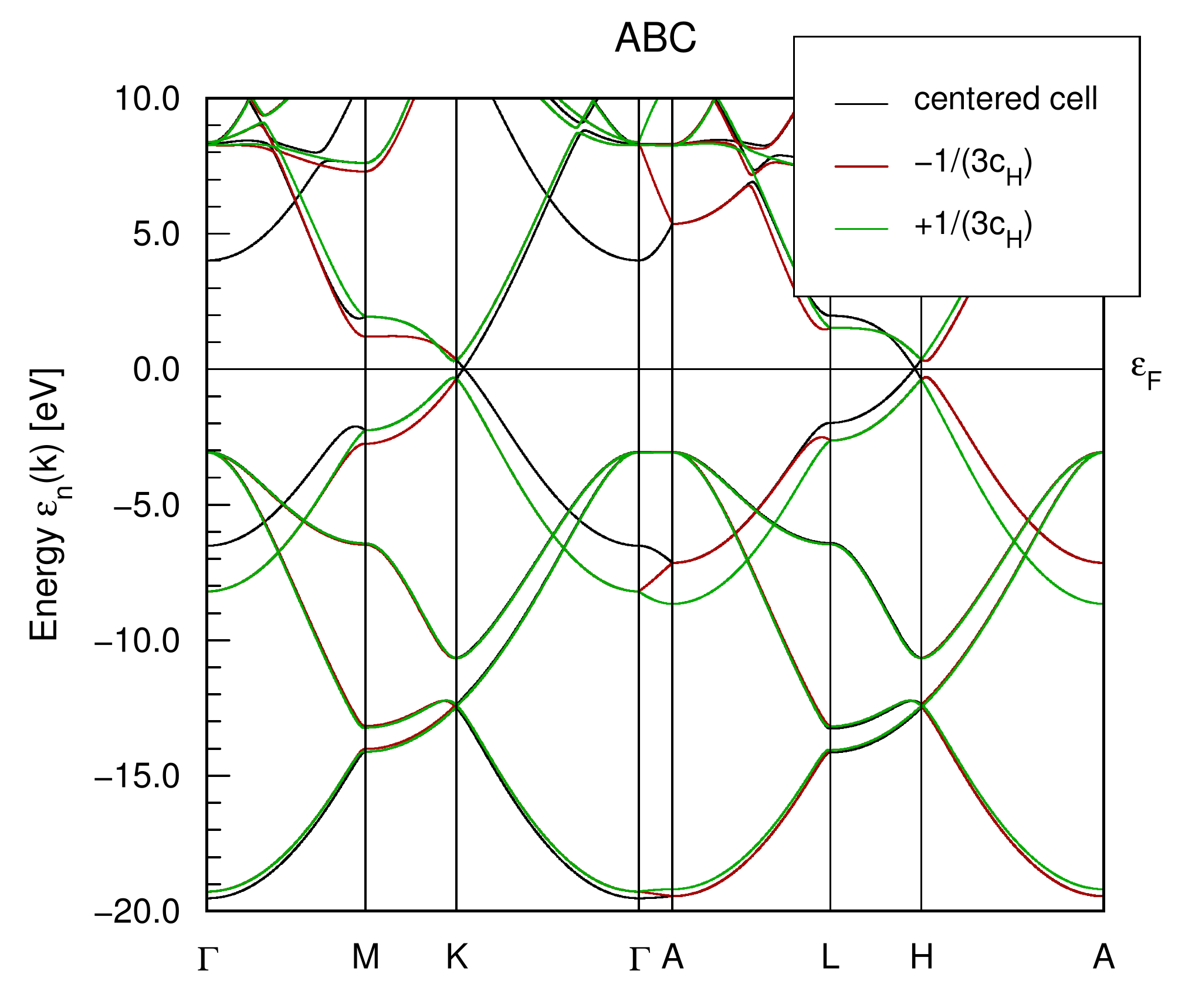}
\par\end{centering}
\caption{\label{fig:hexfromrtg}Band structure calculated in the rhombohedral
2-atom setup. 
Upper panel: Band structure along the lines defined in 
Fig.~\ref{fig:BZ-of-ABC}b;
Lower panel: Band structure along lines of the three hexagonal
BZs shown in Fig.~\ref{fig:BZ-of-ABC}a. Black lines denote bands of
the central BZ, green and red lines denote bands of the upper and
lower BZ, respectively.
This composed band structure
is equivalent to a back-folded band structure obtained in a 6-atom
hexagonal setup using the first hexagonal BZ only.}

\end{figure}

We close this subsection by showing the band structure 
of bulk $\beta$-graphite in a more complete overview.
This aims at clarifying the position of the two Fermi-level
Dirac cones in the rhombohedral trigonal (RTG) BZ
and at comparing with previously published band 
structures~\cite{charlier94}.
In Fig.~\ref{fig:BZ-of-ABC}a the relation between the RTG (2-atom
setup) and the hexagonal (6-atom setup)
BZs is shown. The first BZ of the hexagonal
setup is drawn
opaque in the center of the rhombohedral BZ. The hexagonal
high symmetry points are labelled in white, the RTG points T and X
in black.
The first hexagonal BZ can be shifted by a reciprocal lattice vector
of the hexagonal lattice, which gives the two transparent cells above and
below the opaque cell in the figure 
(shift by $\left(0,0,\pm\frac{1}{3c_{H}}\right)$).
Note, that the X-point (the T-point) of the RTG BZ coincides with the
M-points (the A-points)
of the shifted hexagonal cells. This leads to a back-folding of the
bands around the X-point (T-point) onto the M-point 
(A-point), when the hexagonal 6-atom
setup is used. In Fig.~\ref{fig:hexfromrtg} (lower panel)
the bands obtained from an RTG 2-atom setup along the 
three paths $\Gamma$-M-K-$\Gamma$-A-L-H-A of the
three hexagonal cells are plotted into one figure. 
The resulting band structure is equivalent to the one obtained
from a straight-forward 6-atom hexagonal setup.

Comparing the band structure of Fig.~\ref{fig:hexfromrtg} 
(lower panel) with Fig.~3 of Ref.~\onlinecite{charlier94}
we find that both band structures agree except on the symmetry
points K and H and the related symmetry lines.
It was already noted by the authors of  Ref.~\onlinecite{charlier94}
that the wide-gap insulating behavior found in their hexagonal
setup is not correct. Instead, they find a band crossing
close to the Fermi level on the line L-W in the RTG setup
and an avoided crossing on $\Gamma$-X.

In Fig.~\ref{fig:hexfromrtg} (upper panel)
the band structure as obtained from
the RTG setup along the path shown in Fig.~\ref{fig:BZ-of-ABC}b
is shown. Note, that the part X/m-(U)-k-$\Gamma$ and the part L-h-(W)-T
correspond to the parts M-K-$\Gamma$ and L-H-A of 
Fig.~\ref{fig:hexfromrtg} (lower panel),
respectively. Of course, only the bands belonging to the centered
hexagonal BZ (opaque in Fig.~\ref{fig:BZ-of-ABC}a) are present.
The RTG band structure shown in Fig.~\ref{fig:hexfromrtg} (upper panel)
agrees qualitatively with that published in Ref.~\onlinecite{charlier94},
Fig.~4, with the severe difference that we find a crossing on $\Gamma$-X.
We conclude, that the two Dirac cones close to the Fermi level
of the band structure of bulk (ABC) graphite
shown in Figs.~\ref{fig:bulk+10L}~and~\ref{fig:bulk-projected} 
originate from band crossings on two symmetry lines of
the RTG BZ which coincide with symmetry lines of the 
hexagonal BZ.

\subsection{Topological nature of the surface states}

A remarkable and until now obscured characteristic of the surface states,
present in finite (ABC)-stacked graphene, is their topological nature.
It has been demonstrated in the literature~\cite{guinea06} that a low energy,
long wavelength model (expanded around the K-points of the basal
2D Brillouin zone) for stacked graphene layers can be mapped onto a formally
equivalent one-dimensional (1D) problem,
reminiscent of electrons moving along a 1D chain.
One can define such a mapping for Bernal-stacked graphene as well as
for rhombohedral stacking, resulting in two distinct effective 1D chain models.
When considering two hopping processes~\cite{guinea06},
the tight-binding model for electrons moving along a 1D chain derived
from rhombohedral stacking is in turn equivalent to the famous
Su-Schrieffer-Heeger model for a dimerized chain~\cite{su79,jackiw76}.
The latter model is known to accommodate both trivial and non-trivial
topological phases, depending on the relative strength of the hopping
parameters. The consequence of non-trivial topology is the existence
of zero energy modes, localized on the edges of the 1D chain. 

From this perspective, the surface states in (ABC)-stacked graphene emerge
as a consequence of the non-trivial topology of the bulk model system,
which is revealed by the formal mapping onto a 1D chain.
In order to clarify the topological distinction between rhombohedral
graphite (RG) and Bernal hexagonal graphite (BHG) in the
tight-binding approximation, 
we briefly discuss the 1D models derived from them,
taking into account in-plane nearest neighbor hopping $\gamma_0$
and vertical nearest neighbor interlayer hopping $\gamma_1$.
Both for BHG and RG the mapping onto a 1D chain renders a problem
with a two-site unit cell.
We label the two sites $\alpha$ and $\beta$ and use the wave function basis
$(\psi_{\alpha l}, \psi_{\beta l})^T$,
where $l$ designates the chain unit cell,
$l = 1, \ldots , N$.
The two hopping processes in RG lead to the effective inherited
hopping parameters $v_F |{\bf k}|$ and $\gamma_1$ for the chain.
Here, $v_F = 3a\gamma_0/2$, $a$ denoting the in-plane carbon-carbon distance
and ${\bf k}$ the in-plane momentum in the basal BZ with respect
to the K-point.
In both cases, RG and BHG, we can transform to momentum space.
In terms of the momentum along the chain,
which we call $\theta$ to avoid confusion,
the respective Hamiltonians for RG and BHG read
\begin{equation}
H^{\rm RG}(\theta) = 
(v_F |{\bf k}| + \gamma_1\cos \theta)\; \sigma^x +
 \gamma_1\sin \theta \;\sigma^y
\end{equation}
and
\begin{equation}
H^{\rm BHG}(\theta)  =
 \gamma_1 \cos \theta\;  \sigma^0 +
 \gamma_1\cos \theta   \, \sigma^z +v_F|{\bf k}|\;\sigma^x
\end{equation}
where $\sigma^i$ are the Pauli-matrices ($\sigma^0 \equiv I_2$).
Based on these expressions we can immediately identify how they fit into
the generic form of any 1D two band Hamiltonian, given by 
$H(\theta) =
\varepsilon_0(\theta) \sigma^0+{\bf d}(\theta)\cdot \boldsymbol{\sigma}$.
The topological properties are contained in the vector ${\bf d}(\theta)$,
which reads for RG
${\bf d}(\theta) =
(v_F |{\bf k}| + \gamma_1\cos \theta, \;  \gamma_1\sin \theta, \;  0)$ 
and for BHG
${\bf d}(\theta) = (v_F |{\bf k}|, \;  0, \;\gamma_1\cos \theta)$.
The vector ${\bf d}(\theta)$ may be considered as a mapping from the circle
(due to the periodicity in $\theta$) to the 2D plane~\cite{note1}.
The topological distinction between the two models can now be explicitly
formulated in terms of the winding of the ${\bf d}(\theta)$-vector
around the origin, as $\theta$ is cycled from 0 to $2\pi$.
It is easy to see that for BHG there is no possibility for winding around
the origin, while for RG the ${\bf d}(\theta)$-vector does wind around the
origin if $v_F |{\bf k}| < \gamma_1$.
The latter inequality then is the condition for non-trivial topology and 
hence the presence of edge states (surface states).
It is in full agreement with the results reported in 
Ref.~\onlinecite{guinea06}, but here derived from topological arguments.
From this simplified picture we can gain qualitative insight in the projected
band structure shown in Fig.~\ref{fig:bulk+10L}.
The surface states appear in the vicinity of the K point,
for small momenta $|{\bf k}|$, and have zero energy.
At some point the bulk bands touch after which, for larger $ |{\bf k}|$,
the surface states have disappeared.


The same kind of reasoning can be applied to models that take into
account additional hopping processes.
We have investigated a more involved tight-binding description
of (ABC)-stacked graphene layers,
including hopping processes $\gamma_2$, $\gamma_3$, and $\gamma_4$, 
and found analytically (for $\gamma_3=0$) and numerically
(for $\gamma_3\ne 0$) that the topological distinction is retained.
There remains an area in the in-plane {\bf k}-space where
topological zero-energy modes exist.
We would like to recall, however, that the DFT results
of the previous subsection indicate a bulk (semi)-metallic state
for slabs with $N\ge N^{\rm semimet}$. Thus, the predicted surface
metallic behavior will turn into bulk metallic behavior with 
surface states present in the bulk gap for very thick slabs.

The interpretation of the surface states in terms of topology leads to an
interesting speculation, since zero-energy modes in the 1D dimerized chain are
not only expected on the edges, but also localized at domain walls separating
topologically distinct phases.
In RG one would expect these zero modes to appear when the stacking sequence
is reversed somewhere in the bulk, going from ABCABC to CBACBA.
They would be bound to the layer where the reversal occurs and may have
potentially interesting properties,
different from stacking faults in Bernal stacked graphene.
The same kind of zero-energy modes should be present at boundaries 
between (ABC) and (AB) stackings.

\subsection{Induced charge density and dipole moment }

\begin{figure}
  \centering
  \includegraphics[width=0.80\textwidth]{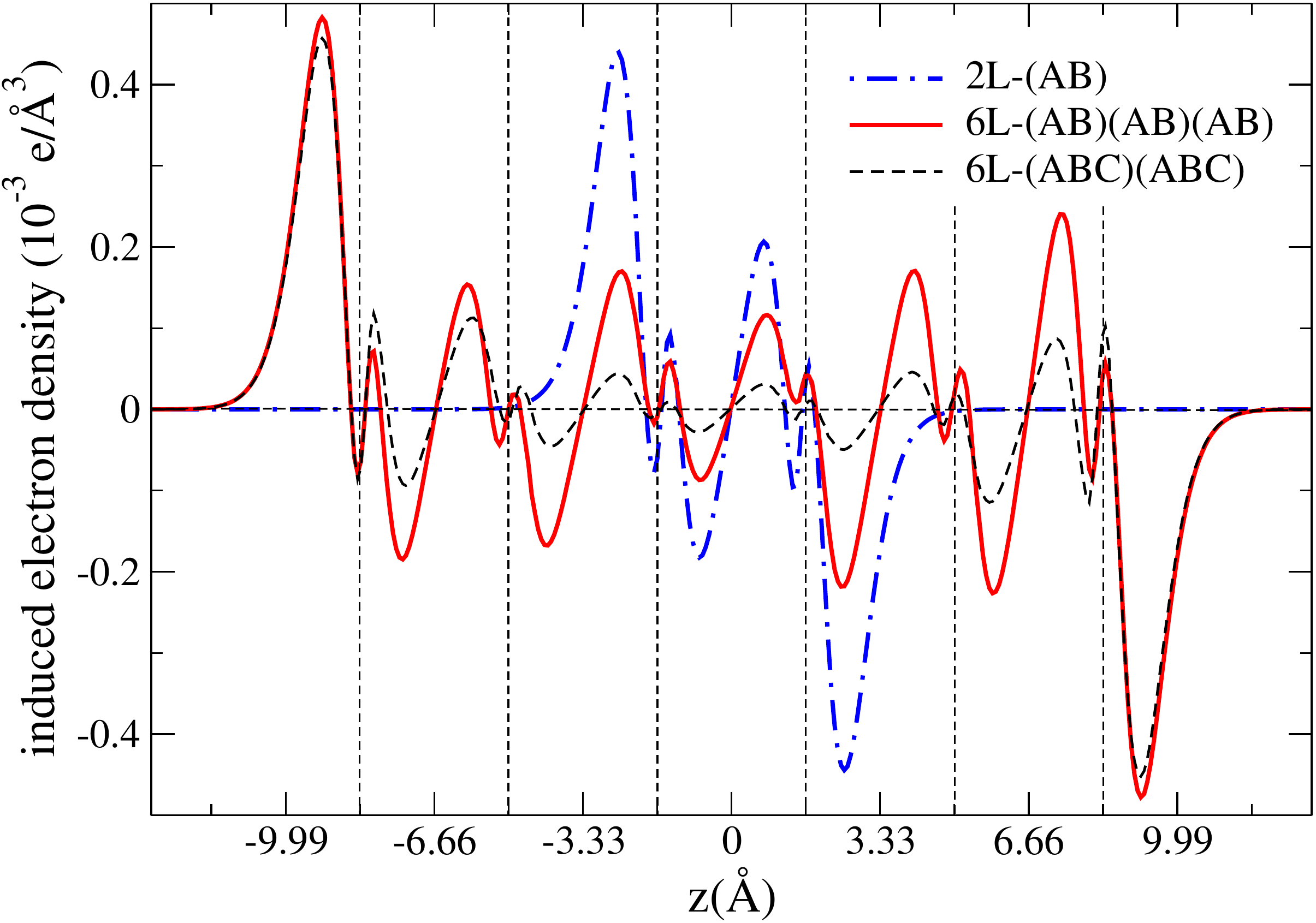}
  \caption{(Color online) Planar average of the induced electron density for an
AB-bilayer compared with 6-layer (AB) and (ABC) stacks
as a function of $z$ (coordinate perpendicular to the slab).
The external field $E_{\rm ext}= 0.1$ V/\AA{} points from the left to the right.
The dashed vertical lines indicate the positions of the carbon planes.
           }
  \label{fig:Induced-CharDens}
\end{figure}
\begin{figure}
  \centering
  \includegraphics[width=0.9\textwidth]{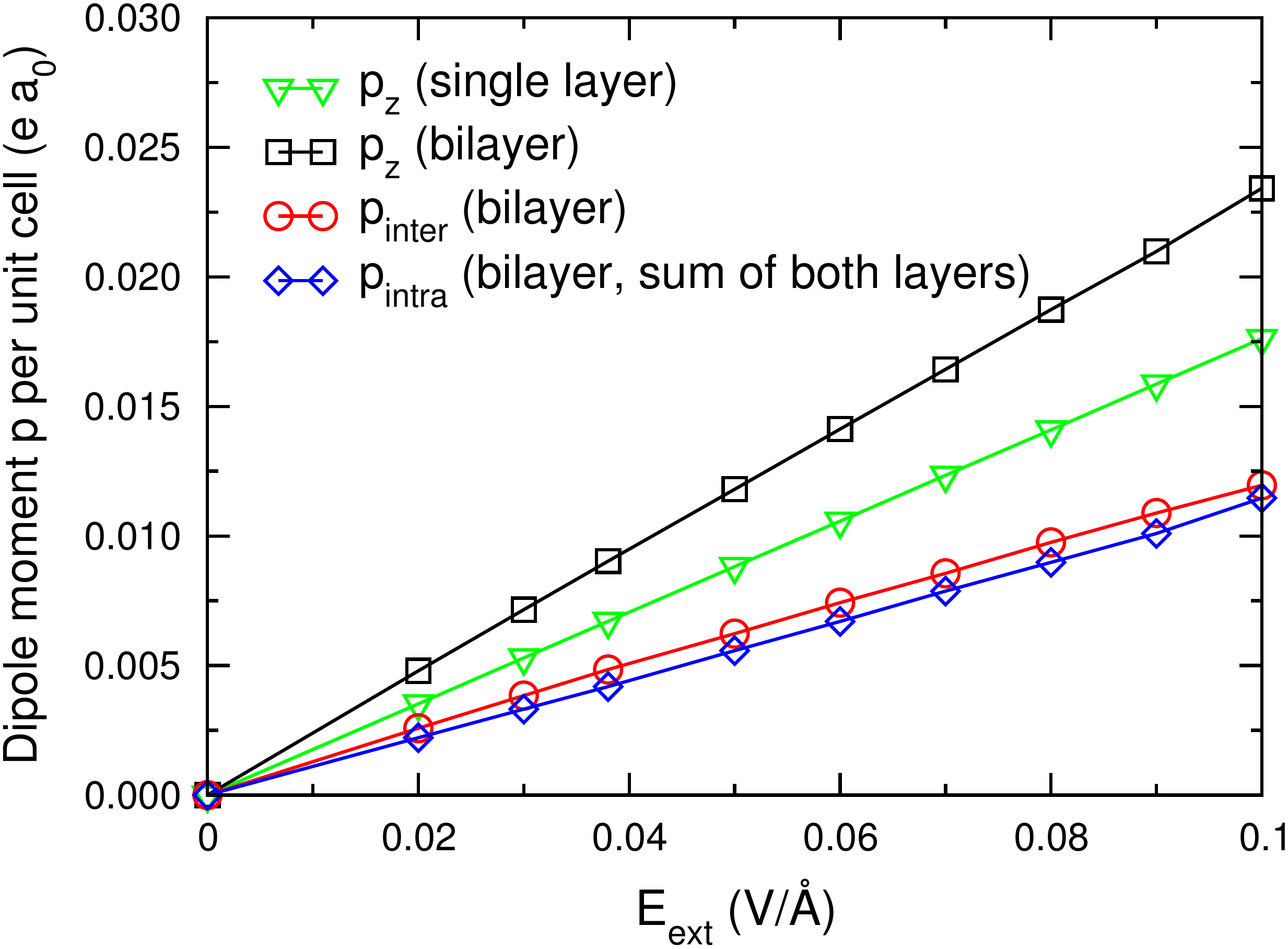}
  \caption{(Color online) Partition of the total dipole moment $p_z$
(given in atomic units e$\cdot$a$_0$ = 8.478$\cdot 10^{-30}$ As$\cdot$m) 
of a bilayer into inter-- and intra--layer contributions.
The total dipole moment of a single layer,
which is only due to intra--layer polarization, is given for comparison.
Lines guide the eye.
           }
  \label{fig:Inter-intra-Dipole}
\end{figure}

We proceed with the discussion of few-layer stacks in an external
electric field along the $z$-direction.
A detailed analysis of the field-induced charge density exhibits some
interesting features which are not shown
by non--selfconsistent tight binding models.
We consider an average of the induced charge density in the $x-y$ plane,
$\bar{n}(z)$, which can be used to evaluate the induced dipole moment
per unit cell $p_z$,
\begin{equation}
p_z = -|{\rm e}| A_{\rm uc} \int dz \;z \;\bar{n}(z) \; ,
\end{equation}
where ${\rm e}$ denotes the elementary charge and $A_{\rm uc}$
the area of the unit cell in the $x-y$ plane.

The behavior of $\bar{n}(z)$ presented in Fig.~\ref{fig:Induced-CharDens}
shows a significant asymmetry around each carbon plane.
Thus, it makes sense to distinguish an intra-layer and an inter-layer
contribution to the dipole moment, $p_z = p_{\rm intra} + p_{\rm inter} $.
We intend to analyze these contributions only for the bilayer case
and define $p_{\rm inter} = Q d$. Here, $Q$ denotes the induced charge
per layer, obtained as sum of projections to the two atomic sites,
and $d=3.33$ \AA{} is the layer distance.
Fig.~\ref{fig:Inter-intra-Dipole} shows that in the bilayer both contributions 
have approximately equal size. 
Both contributions produce a dipole moment in the same direction which is consistent with 
the classical notion of a density shift in the direction of the Coulomb force 
due to the external field.
Fig.~\ref{fig:Inter-intra-Dipole} includes also the dipole moment
calculated for a single layer (graphene),
which is about three times larger than the {\em intra}--layer contribution
of one layer in a bilayer system (observe, that the related curve
in Fig.~\ref{fig:Inter-intra-Dipole} shows the sum of both
{\em intra}--layer contributions in the bilayer system).
The reason for this difference is that in a mono-layer system there is no
inter--layer charge transfer that screens an 
essential part of the external field.

Fig.~\ref{fig:Induced-CharDens} shows
the planar average of the induced electron densities for an AB-bilayer
and for 6-layer systems with (AB) and with (ABC) stackings.
Due to the screening of the external field the 
amplitudes of the induced electron density 
are significantly smaller in the interior of the 
6-layer slabs than on their surface. 
This feature is in contrast to the results for (AB) stacks
published in Ref.~\onlinecite{yu08}, 
where the amplitudes of the induced electron densities 
in the middle of the 6-layer system differs from the surface layer
amplitudes by a few percent only. 
These results however refer to electric fields which are one
order of magnitude weaker than in our calculation.
In order to ensure reliability of our data,
we repeated the FPLO calculations with the QE code and
found virtually the same results.
Note, that both codes are completely independent.
Not understandable from a fundamental point of view are
induced electron densities published in Ref.~\onlinecite{yu08}
for bulk graphite.
Though $\alpha$-graphite is metallic and should not show any
induced charges in the bulk, the induced bulk charge
densities shown in Fig.~3(d) of Ref.~\onlinecite{yu08}
are yet larger than the surface charge densities of the
6-layer slab.

It is interesting to note,
that the ABABAB slabs show a larger induced screening
charge density in the inner layers than the ABCABC slabs.
This seems to contradict the fact, that ABABAB slabs stay
metallic for the electric field range in question,
whereas the ABCABC slabs turn semi-conducting. 
Qualitatively, the same trend is seen in the results from 
the simplest possible tight binding model 
calculation~\cite{koshino10}.
We suggest that the observed astonishing
behavior can be understood from the
different character of states contributing to the screening.
In rhombohedral stacks, only the two surface states 
are close to the Fermi level. Thus, the screening charge
density is
concentrated close to the surface. In Bernal stacks,
${\cal O}(N)$ states are present at the Fermi level
in the absence of an external field. Hence, screening
charge density is contributed from a number of layers
if a field is applied.
Obviously, the hand-waving argument that screening in metals 
is more effective than in semi-conductors 
does not necessarily apply to few-layer slabs, where the
asymptotic exponential behavior is not yet valid.
The following analysis will however show,
that the layer-integrated charges decay on a shorter
length-scale than the charge densities.

\subsection{ Inter-atomic charge transfer}

\begin{figure}
  \centering
\begin{minipage}{.5\textwidth}
{\large chemical electron transfer $N_{chem}$}
\end{minipage}\\[.5cm]
\begin{minipage}{.43\textwidth}
  \includegraphics[width=0.7\textwidth]{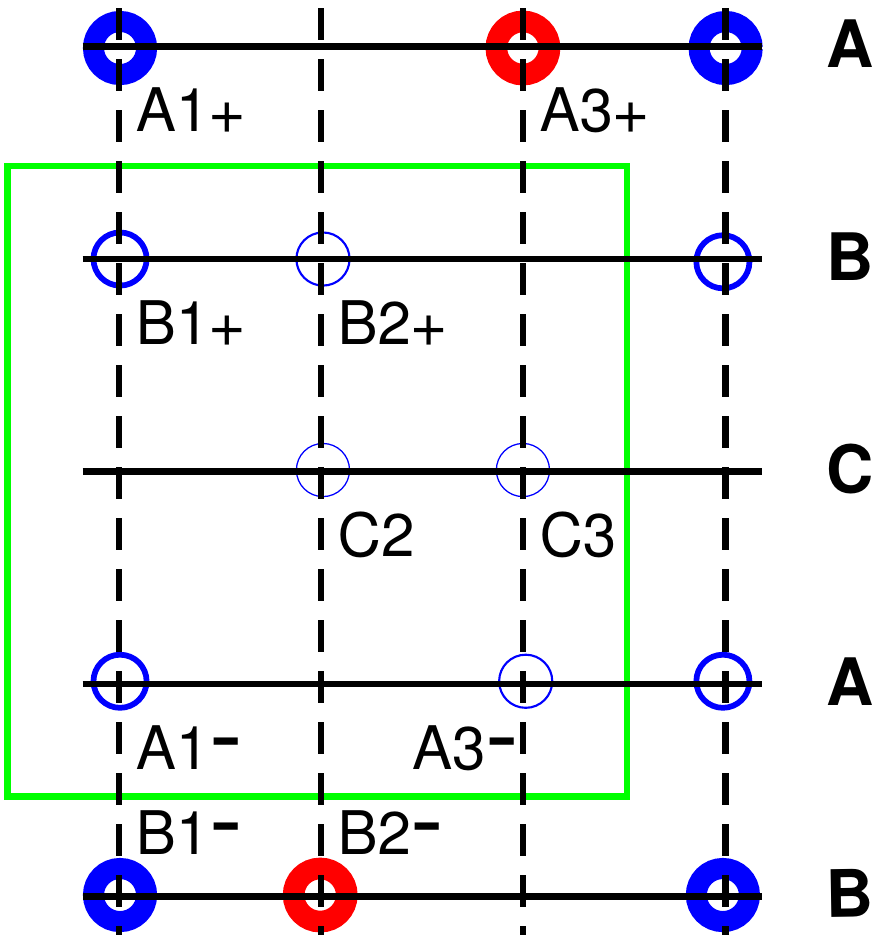}
\end{minipage}
\begin{minipage}{.43\textwidth}
  \includegraphics[width=0.7\textwidth]{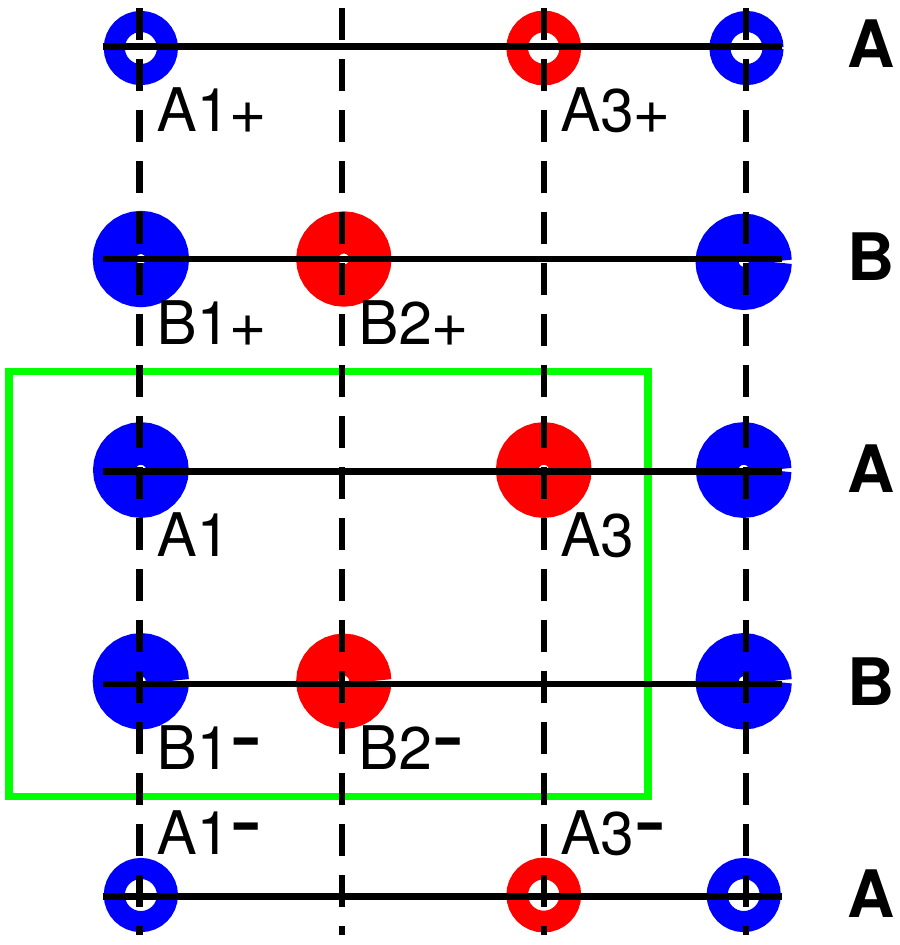}
\end{minipage}\\[.5cm]
\begin{minipage}{.5\textwidth}
{ \large physical electron transfer $\Delta N_{phys}$}
\end{minipage}\\[.5cm]
\begin{minipage}{.43\textwidth}
  \includegraphics[width=0.7\textwidth]{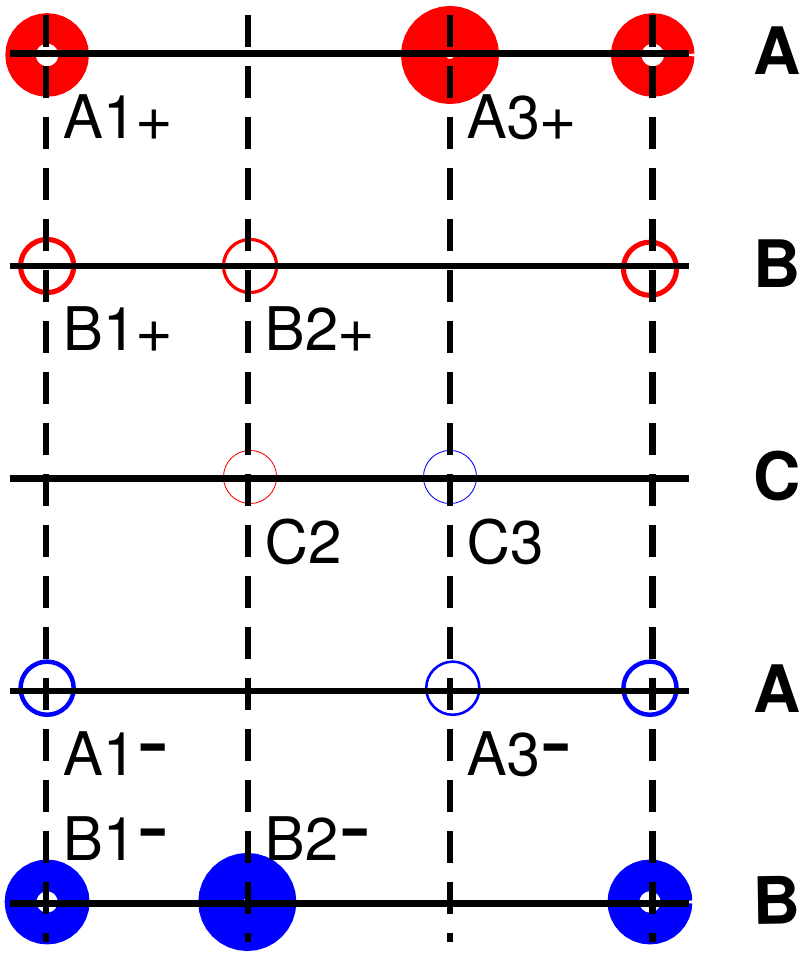}
\end{minipage}
\begin{minipage}{.43\textwidth}
  \includegraphics[width=0.7\textwidth]{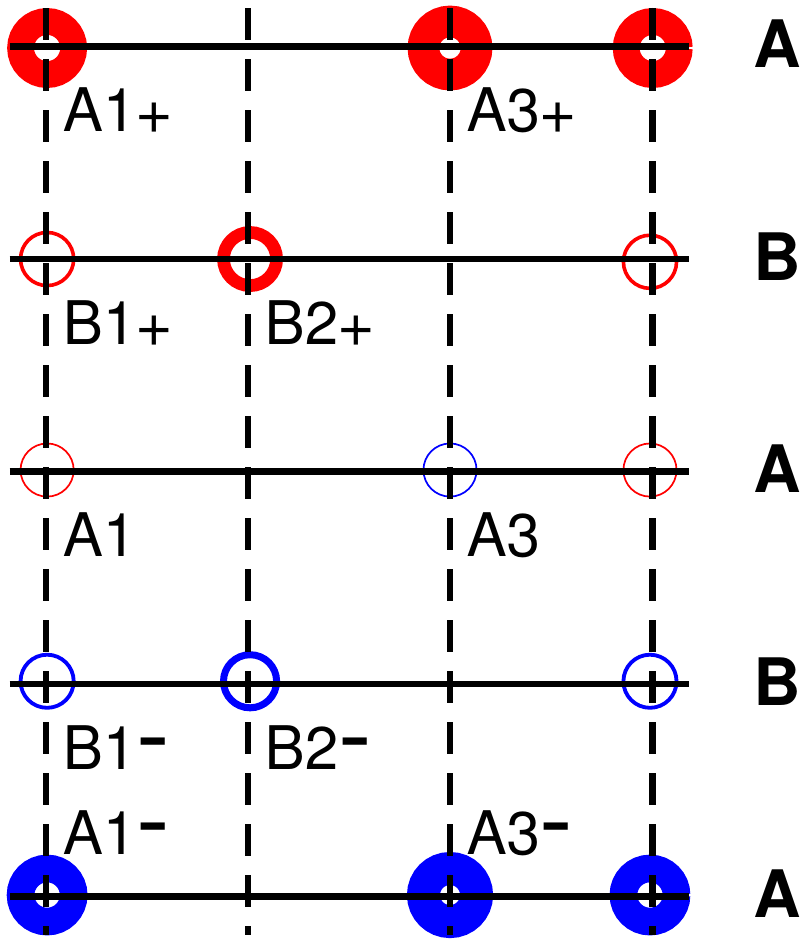}
\end{minipage}\\[.5cm]

  \caption{ (Color online)
{\em Chemical} electron transfer $N_{chem}$ (upper panels)
and {\em physical} electron transfer $\Delta N_{phys}$ (lower panels)
in  5 layer graphene slabs with
(ABC) stacking (left panels) and (AB) stacking (right panels).
The excess charge is proportional
to the line width of the circles.
For better graphical presentation,
the physical transfer is enlarged by a factor of 10
compared with the chemical transfer.
Red (light) and blue (dark) stands for positive and negative values, respectively.
Stacking direction is from top to bottom. The electric field 
of a strength $0.1$ V/\AA{} points downward
(external force on the electrons
upward). The signs in the notation of A and B layer atoms refer
to their location above or below the central layer. The numbers denote 
the position
of the projection  of the atomic position in the 2D unit cell.
The green boxes mark the repetition unit in the slabs.
}
  \label{fig:transfer}
\end{figure}

Table~\ref{table:excess-electrons} and Fig.~\ref{fig:transfer} show the
electron transfer between the atoms for both (ABC) and (AB) stacking,
calculated by atomic site projections.

{\em Without external field} (see upper panel of Fig.~\ref{fig:transfer}
and left columns of Table~\ref{table:excess-electrons})
the layers are not completely neutral
(except in the bilayer case), but there is a small electron
transfer toward the surface so that the surface layer
is slightly negatively charged.
The electron flow from the interior to the surface produces
a surface dipole barrier and decreases
the ionization energy or work function.
Additionally, there is a much stronger {\em chemical} {\em intra}-layer
transfer away from
saturated $p_z$ bonds towards dangling bonds
(e.g. from position '1' to '2' or '3' in the surface layers
in Fig.~\ref{fig:transfer}).
For (ABC) stacking, this transfer converges to zero toward the
interior, because in the bulk all six lattice positions
in the repetition unit are physically equivalent  for symmetry reasons.
(Note that the bulk unit cell for (ABC) material is different
from this repetition unit: 
it contains only two atoms and has rhombohedral symmetry.)
For (AB) stacking, however, the intra-layer electron transfer
converges to a finite value, because there are two different
lattice positions in the unit cell:
one has only saturated and one has only dangling bonds,
Fig.~\ref{fig:ab_vs_abc}.

\begin{table}[tbh]
\begin{tabular}{l|cc|cc|cc|cc}
\hline
\multirow{2}{*}{(ABC)}&\multicolumn{2}{|c|}{2L}&\multicolumn{2}{|c|}{3L}&\multicolumn{2}{|c|}{4L}&\multicolumn{2}{|c}{5L}\\
\cline{2-3}\cline{4-5}\cline{6-7}\cline{8-9}
&$N_{chem}$&$\Delta N_{phys}$&$N_{chem}$&$\Delta N_{phys}$&$N_{chem}$&$\Delta N_{phys}$&$N_{chem}$&$\Delta N_{phys}$\\
\hline
A1+&$-$6.74&+0.79&$-$6.58&+0.90&$-$6.56&+0.94&$-$6.55&+0.95\\
A3+&+6.74&+0.97&+6.79&+1.26&+6.82&+1.37&+6.84&+1.41\\
\hline
B1+&$-$6.74&$-$0.80&$-$0.22& 0.00&$-$0.18&+0.10&$-$0.18&+0.16\\
B2+&+6.74&$-$0.96&$-$0.22& 0.00&$-$0.07&+0.07&$-$0.07&+0.10\\
\hline
C2&&&$-$6.58&$-$0.91&$-$0.07&$-$0.07&$-$0.04&+0.03\\
C3&&&+6.79&$-$1.23&$-$0.18&$-$0.11&$-$0.04&$-$0.02\\
\hline
A1$-$&&&&&+6.82&$-$1.35&$-$0.18&$-$0.14\\
A3$-$&&&&&$-$6.56&$-$0.96&$-$0.07&$-$0.08\\
\hline
B1$-$&&&&&&&$-$6.55&$-$0.99\\
B2$-$&&&&&&&+6.84&$-$1.41\\
\hline
\hline
\multirow{2}{*}{(AB)}&\multicolumn{2}{|c|}{2L}&\multicolumn{2}{|c|}{3L}&\multicolumn{2}{|c|}{4L}&\multicolumn{2}{|c}{5L}\\
\cline{2-3}\cline{4-5}\cline{6-7}\cline{8-9}
&$N_{chem}$&$\Delta N_{phys}$&$N_{chem}$&$\Delta N_{phys}$&$N_{chem}$&$\Delta N_{phys}$&$N_{chem}$&$\Delta N_{phys}$\\
\hline
A1+&$-$6.74&+0.79&$-$6.56&+0.79&$-$6.58&+0.81&$-$6.57&+0.84\\
A3+&+6.74&+0.97&+6.80&+0.89&+6.81&+0.95&+6.81&+0.99\\
\hline
B1+&$-$6.74&$-$0.80&$-$13.63&$-$0.05&$-$13.47&+0.06&$-$13.49&+0.11\\
B2+&+6.74&$-$0.96&+13.15&$-$0.05&+13.25&+0.27&+13.23&+0.40\\
\hline
A1&&&$-$6.56&$-$0.75&$-$13.47&$-$0.08&$-$13.30&$-$0.05\\
A3&&&+6.80&$-$0.82&+13.25&$-$0.14&+13.34&$-$0.05\\
\hline
B1$-$&&&&&$-$6.58&$-$0.86&$-$13.49&$-$0.12\\
B2$-$&&&&&+6.81&$-$1.02&+13.23&$-$0.22\\
\hline
A1$-$&&&&&&&$-$6.57&$-$0.87\\
A3$-$&&&&&&&+6.81&$-$1.03\\
\end{tabular}

\caption{Number of excess electrons (in units $10^{-3}$)
for atoms in $2 \ldots 5$-layer graphene slabs
without external field ($N_{chem}$) and the number of
electrons ($\Delta N_{phys}$) induced by an
external electric field of $0.1$ V/\AA{}.
The external electric field points downwards (external force
on the electrons upwards).
The upper and lower panel contain data for (ABC) and (AB) stacking,
respectively.}
\label{table:excess-electrons}
\end{table}

{\em The external field}  produces an average
{\em physical} {\em inter}--layer transfer
in the direction of the external force 
from one surface to the other, whereby  
the induced excess electrons (holes) are  
well localized in the very surface layer (see lower panel
of Fig.~\ref{fig:transfer} and right columns
of Table~\ref{table:excess-electrons}).
Moreover, the induced excess charge is almost equally 
spread over both atoms within the surface layers.
This is not in contradiction to Fig.~\ref{fig:Inter-intra-Dipole}
which shows a sizable $z$-dependence of the induced electron density 
even in the interior, as the
$z$-dependence of the induced density in the interior 
is mainly due to intra-layer polarization, which contributes to the 
dipole moment, but not to the total excess charge.
We also observe that 
the positively and the negatively charged {\em surface} layers
(i.e. the lower versus the upper layers) do not carry the same
absolute amount of induced charge, except for the bilayer.
In other words, the absolute excess charge is not symmetric
in $z$-direction. 
As an example, in the (ABC) 5-layer system
the upper (negative) surface layer has
$+2.36 \cdot 10^{-3}$ induced excess electrons per layer cell,
whereas the lower positive) surface layer has
$-2.40 \cdot 10^{-3}$.
Reasons for this asymmetry 
can be the lacking mirror symmetry of the
(ABC) system and/or lacking electron/hole symmetry of
the electronic structure around the Fermi level.
Since the charge transfer is yet more asymmetric in the
(AB) systems, the latter reason is probably more
important.

We also want to draw the attention to the fact 
that even for the maximum considered
external field strength 
the {\em chemical intra}--layer transfer 
is about one order of magnitude larger than the 
{\em physical inter}--layer transfer.

\subsection{Consequences for modelling the dielectric response 
of graphene stacks}

What can we learn from the results 
in the previous subsections about the validity of 
simple models  and the notions of macroscopic electrodynamics 
for the dielectric response in graphene stacks 
with microscopic thickness?

The fact that almost all {\em induced charge} per layer 
is located within the very 
surface layer is consistent with the notion of location of the induced 
charge on the mathematical surface in the model for metals in 
 macroscopic electrodynamics.
This trait is also seen in the results of the 
simplified screening models \cite{
mccann06,      
mcdonald10a}   
applied to stacks of up to 20 layers 
\cite{koshino10}, 
which treat the atomic layers as mathematical planes carrying the 
screening charge.  

On the other hand, as seen in the induced charge densities
in the last but one subsection, the {\em induced dipole moments} per layer,
which drop off very slowly in moving inward, 
call for a model which is reminiscent of the 
picture with localized polarizable dipoles traditionally used for insulators. 
This contribution to the screening field is not included in the 
simplified models \cite{mccann06,mcdonald10a,koshino10} at all.
Consequently, these models can provide only qualitative answers, and 
the graphene stacks behave neither like traditional metals nor like 
traditional insulators.

\subsection{Local density of states (LDOS) }

\begin{figure}
  \centering
  \includegraphics[width=0.8\textwidth]{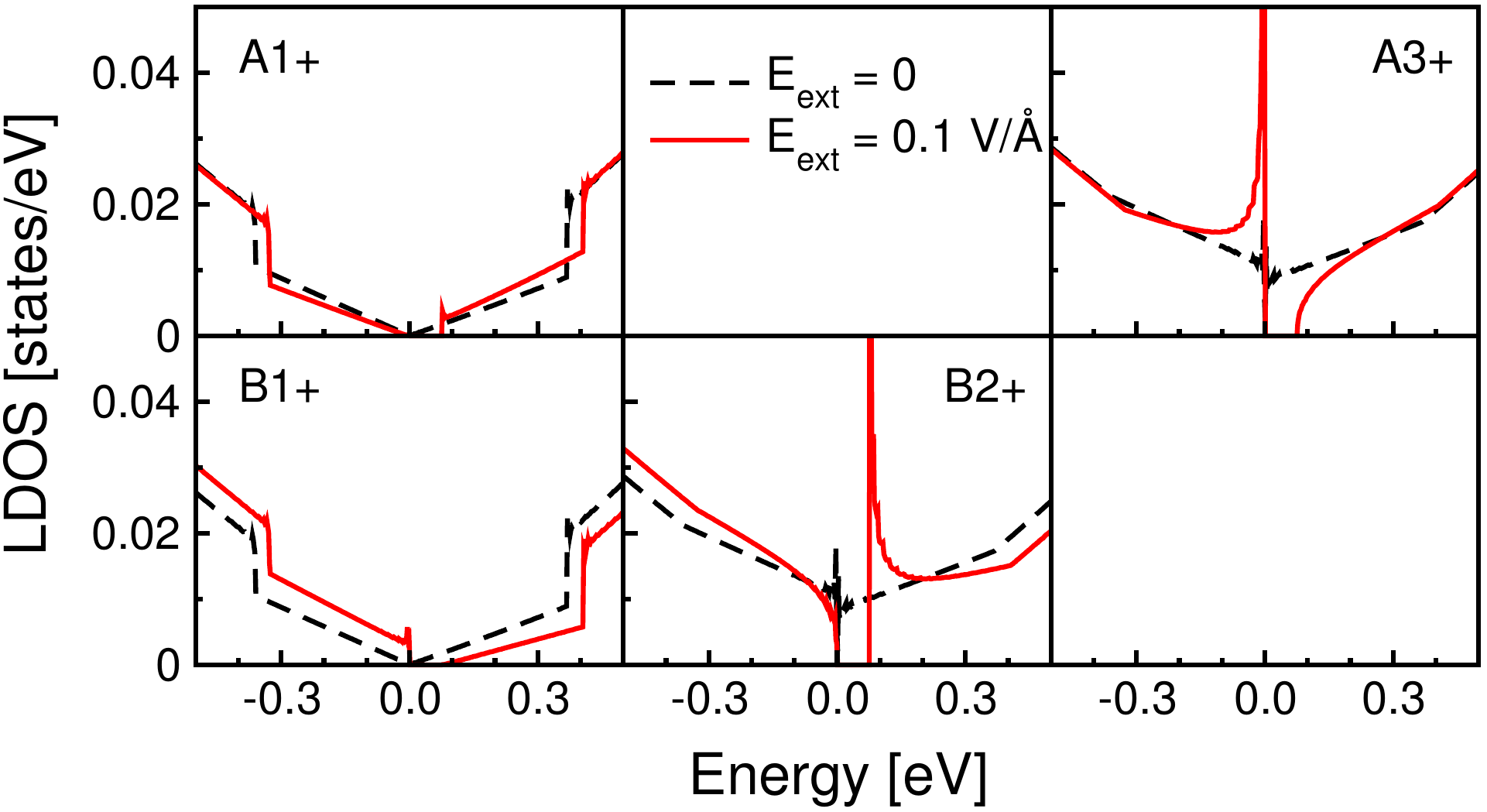}
  \caption{ (Color online)
LDOS for the $p_z$ states of AB-bilayer atoms for vanishing 
and for finite external field, $E_{\rm ext}=0.1$ V/\AA.
The panels are arranged according to the atomic positions
in the notation used in Fig.~\ref{fig:transfer}.
Despite a dense mesh of points in $\bf k$ space with $600 \times 600$ points in
the 2D Brillouin zone
and a linear interpolation of the energies and matrix elements within triangles,
the visible numerical fluctuations could not be avoided.
The Fermi level is at zero energy and the gap at the finite field
amounts to 0.074 eV.
The energy grid spacing is 0.002 eV.
           }
  \label{fig:LDOS-2L}
\end{figure}

\begin{figure}
  \centering
  \includegraphics[width=0.8\textwidth]{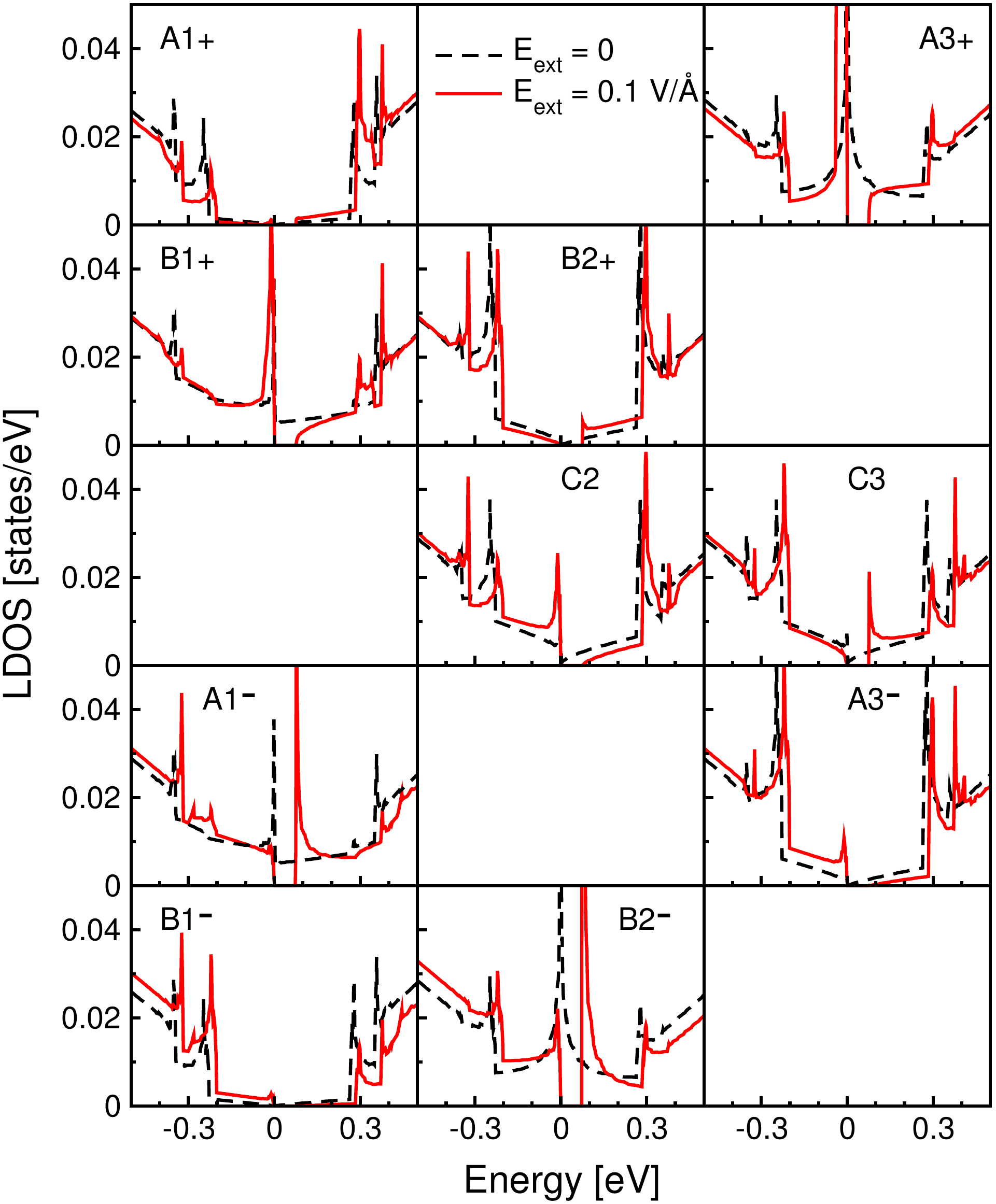}
  \caption{ (Color online)
LDOS for the $p_z$ states of a 
5-layer (ABC) slab for vanishing 
and for finite external field, $E_{\rm ext}=0.1$ V/\AA.
The panels are arranged in analogy to the atomic 
pattern in Fig.~\ref{fig:transfer}.
The Fermi level is at zero energy
and the gap at the finite field 
amounts to 0.072 eV. The energy grid spacing is 0.002 eV.
           }
  \label{fig:LDOS-5L}
\end{figure}

Fig.~\ref{fig:LDOS-2L} shows the LDOS for the bilayer near the Fermi energy.
The chemical charge transfer (for vanishing field)
from atoms A1+ and B1+ to A3+ and B2+ 
(see Fig.~\ref{fig:transfer} and Table~\ref{table:excess-electrons})
is reflected in the fact that  the LDOS  below the Fermi level for the atoms
with excess charge (A3+ and B2+)
is much higher than for the deficit atoms (A1+ and B1+). 
An electric field changes the analytical 
properties of the energy bands qualitatively: the {\em cusps} turn 
into {\em Mexican hat} structures, Fig.~\ref{fig:BandStr}.
Because the iso-energy lines of the extrema are 
roughly speaking circles, which resembles 
a one-dimensional band structure, 
the total DOS (sum of all LDOS) must exhibit $1/\sqrt \varepsilon$-type 
singularities at the edges of the gap \cite{guinea06}.
The intuitive guess that
the singularity at the occupied side of the 
gap should appear in the LDOS of atom A3+ and
the singularity at the unoccupied side of the gap on atom B2+
is confirmed by the calculation.
Both atoms have dangling $p_z$-bonds toward the interior of the slab. 
The occupied peak at A3+ is connected with the 
physical charge accumulation at the A3+ atom
(see Table~\ref{table:excess-electrons}) in the electric field.

Fig.~\ref{fig:LDOS-5L} shows the LDOS of the ABCAB slab.
The LDOS of the surface atoms of this stack is expected to 
show virtually the features of an infinitely thick slab. 
Despite the fact that Fig.~\ref{fig:LDOS-5L}
shows much more singularities and structures than the bilayer  due to the 
more complicated  band structure, the LDOS
of the surface atoms {\em near the Fermi energy} in the electric field 
is very similar to the bilayer: 
the LDOS at the band edges
of the surface atoms with dangling bonds have singularities
and the LDOS of the other surface atoms is very small.
This is partly due to the fact that 
the energy bands in the external field 
in the vicinity of the Fermi energy
shown in Fig.~\ref{fig:BandStr} are similar: all 
exhibit a Mexican hat shape.
These singularities at the band edges are still 
observable in a weaker form at atoms with inward pointing 
dangling bonds, 
which are located in the second layer below the surfaces (B1+ and A1-). 
As a rule, 
the singularities are located on the occupied side
 of the gap in the negatively  charged layers 
and on the unoccupied side of the gap for 
positively charges layers (see  Figs.~\ref{fig:transfer} 
and \ref{fig:LDOS-5L}).
As shown in \cite{mak10} for the 4-layer system, 
the corresponding singularities in the joint density of states 
are clearly seen in infra-red absorption spectra and can be used 
to identify the stacking order of stacks prepared by exfoliation. 


\subsection{Dielectric constant}

\begin{figure}
  \centering
  \includegraphics[width=0.8\textwidth]{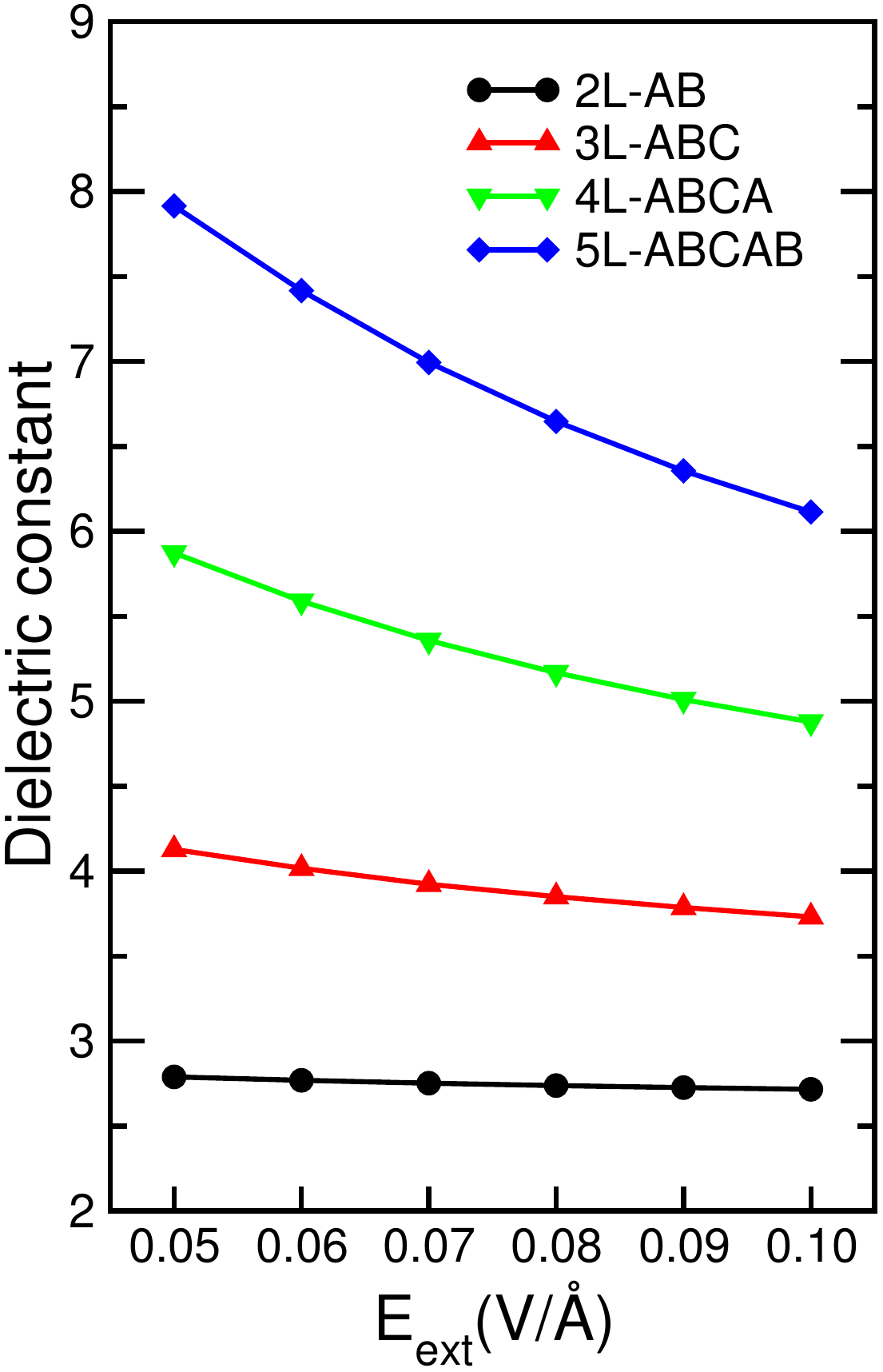}
  \caption{(Color online)
Averaged nonlinear dielectric constant for the AB-bilayer 
and for $3 \ldots 5$-layer
(ABC)-type graphene systems versus $E_{\rm ext}$.}
  \label{fig:EpsvsE}
\end{figure}

We are aware of the fact that a general response relation in 
inhomogeneous systems should include nonlocal effects,
\begin{equation}
E_{\rm int}({\bf r}) =
\int d{\bf r'} \; \epsilon^{-1}({\bf r},{\bf r'}) \; E_{\rm ext}({\bf r'}) \; .
\label{dep-epsilon-nonlocal}
\end{equation}
Therefore, for the extraction of the full information about
the dielectric response function $\epsilon^{-1}({\bf r},{\bf r'})$ 
from a finite field calculation
we had to apply  $\bf r'$-dependent external fields $E_{\rm ext}({\bf r'})$. 
The local field corrections included in Eq.~(\ref{dep-epsilon-nonlocal})
imply that even 
for a constant external field the internal field $E_{\rm int}$
is $\bf r$-dependent. 
Because it is numerically extremely difficult to extract the full information 
about $\epsilon^{-1}({\bf r},{\bf r'})$ from a finite field approach, we 
here use a mesoscopic description by
a dielectric constant (DC) $\epsilon$
which is independent of $\bf r,r'$ within the slab. 
To this end, we follow macroscopic electrodynamics and determine
a spatially averaged internal field $E^{\rm av}_{\rm int}$ from
\begin{equation}
E^{\rm av}_{\rm int}=E_{\rm ext}-4 \pi P(E_{\rm ext}) \; ,
\label{eq:eav}
\end{equation}
where $P$ is the dipole moment per volume, induced by the homogeneous
external field $E_{\rm ext}$. 
For the definition of the volume of the $N$-layer slab we used a thickness
of $N\cdot d$.
Thus, $P=p_z/(A_{\rm uc}Nd)$.
We consider only the $z$-components 
of $P$, $E_{\rm ext}$, and $E_{\rm int}^{\rm av}$,
since by symmetry the other components are zero.

The next point to note is that the gap was found to be strongly
dependent on the strength of $E_{\rm ext}$, see Fig.~\ref{fig:GapvsE}.
This implies that also $\epsilon$ should be strongly field dependent.
Thus, we take a differential definition,
\begin{equation}
d E_{\rm int}^{\rm av}(E_{\rm ext}) =
\epsilon^{-1}(E_{\rm ext})\cdot  d E_{\rm ext} \; .
\label{def-epsilon-lin}
\end{equation}
The more commonly used alternative of a power expansion of
$E_{\rm int}$ in $E_{\rm ext}$ with
field-{\em independent} expansion coefficients is not 
suited for strong nonlinearities,
in particular in cases like ours where the gap depends strongly on the
external field and for vanishing field 
the system is (semi)metallic.

In order to establish a relation with Eq.~(\ref{eq:eav}),
we define an averaged DC $\bar{\epsilon}$ via integration 
of Eq.~(\ref{def-epsilon-lin}),
\begin{equation}
\frac{1}{E_{\rm ext}}
\int_0^{E_{\rm ext}} d E'_{\rm ext}\cdot \epsilon^{-1}(E'_{\rm ext})
= E_{\rm int}^{\rm av}(E_{\rm ext}) / E_{\rm ext} \stackrel{\mathrm def}{=}
\bar{\epsilon}\,^{-1}(E_{\rm ext}) \; .
\end{equation}
Using Eq.~(\ref{eq:eav}), we find
\begin{equation}
\bar{\epsilon}\,^{-1}(E_{\rm ext})=1-4\pi \frac{P(E_{\rm ext})}{E_{\rm ext}} \; .
\end{equation}
Despite the somewhat loose connection to the fully non-local 
dielectric response function, 
$\bar{\epsilon}$ provides exact information about the induced dipole moment per 
surface area.

As seen in Fig.~\ref{fig:EpsvsE},
$\bar{\epsilon}(E_{\rm ext})$ decreases with increasing $E_{\rm ext}$. 
This trend can be understood from the fact that the gap width grows
monotonously with the external field in the considered field range,
Fig.~\ref{fig:GapvsE}.
Another trend also seen in Fig.~\ref{fig:EpsvsE}
is the increase of $\bar{\epsilon}$  
with increasing number of layers $N$ for fixed external field.
For $N>2$ this trend 
can be qualitatively explained with the same argument as before,
since the gap width decreases with increasing $N$. 
Obviously the opening of a gap by an external field 
is the harder the thicker the slab is.   
The bilayer $N=2$ is an exception, because in this system
there is no next-neighbor inter--layer 
interaction.

An interesting application is offered by the distinct values of
$\bar{\epsilon}$ for a relatively large range of external field
values. Namely, a measurement of $\bar{\epsilon}$ should allow to
discriminate the number of layers and, perhaps, also the stacking type.
Such a discrimination was only recently achieved by synchrotron-based
infrared absorption-spectroscopy~\cite{mak10}.

We should finally mention the small absolute value for the DC 
of the bilayer system,
about 2.7 in the considered field range. This is well below 
the value for $SiO_2$, $\epsilon=3.9$, and might be relevant for 
fast nano- or molecular electronics.

\section{Summary and Conclusions}

We performed a DFT investigation of the electronic structure and of
the electric-field response of (ABC)-stacked graphene slabs with
finite thickness. For comparison, related calculations were carried
out for single layer graphene, AB-bilayer, 5-layer and 6-layer
(AB)-slabs. Further, the electronic structure of bulk $\beta$-graphite
was clarified.
The main results can be summarized as follows:\\
(i) The ABC-trilayer is a zero-gap semi-conductor,
but all other (ABC) slabs up to 10
layers thickness and also bulk (ABC) graphite ($\beta$-graphite)
are semi-metals.
Thus, very probably {\em all} (ABC)-slabs with more than 3 layers
are semi-metallic.\\
(ii) We confirmed that in (ABC) slabs up to at least 10
layers thickness
only surface states are located at the Fermi energy
and provided a topological argument based on a tight-binding Hamiltonian
that this should hold for layer thickness larger than 10 as well.
This means, that electronic transport parallel to 
such slabs is confined to a surface region. 
Another, more general implication arising from our
topological argument is that
conducting zero-energy modes also should appear at
grain boundaries between (AB) and (ABC) stacking
and at domain walls separating (ABC) from (CBA)
regions. Since finite-thickness (ABC) stacking is
common both in natural and in synthesized graphite, the electric
conductance of graphite should be very sensitive to
the specific amount of stacking domain walls 
and grain boundaries.\\
(iii) The band structure of the AB-bilayer has been thoroughly
revisited. We found on the meV-scale a tiny electron pocket around
the K-point and a hole pocket on the line K-$\Gamma$ close to K.
Thus, already an undoped AB-bilayer has a finite Fermi line,
in contrast to results from tight-binding calculations with
up to three hopping parameters and in agreement with other
DFT results.
We found that meV-features and the topology of
the Fermi surfaces are very delicate.
Thus, tight-binding results using overlap parameters from
bulk graphite may on the low-energy scale considerably differ
from self-consistent DFT results.\\
(iv) Band structures of several few-layer (ABC) slabs
and of bulk (ABC) graphite were presented and critically
compared with available literature data.
The low-energy band structure of bulk $\beta$-graphite
consists of two Dirac-like cones above and
below the Fermi level with an energy distance of 9 meV.
Since for few-layer (ABC) slabs there is a gap
in the bulk-like states, we conclude that 
there is a critical thickness $N^{\rm semimet} \gg 10$
beyond which (ABC) slabs are {\em bulk} semi-metallic.\\
(v) (ABC)-stacked graphene systems with $3\ldots 5$
layers were found to open a band gap in external electric fields,  
which increases with the electric field strength
in the considered range up to 0.1 V/\AA. \\
(vi) The strong electric-field dependence of the
electronic structure of (ABC) slabs yields a related field dependence 
of the dielectric response. A properly defined field-averaged
dielectric constant $\bar{\epsilon}$ decreases with increasing
external field.
Both the band gap and the dielectric constant are
determined by the number of graphene layers and
systems with different numbers of layers show different
tuning ratios $d\bar{\epsilon}/dE_{\rm ext}$ for the
dielectric constant.
These properties provide a chance to choose a specific
graphene system according to application requirements. 
Vice versa, the specific dielectric response can be
used to identify the number of layers in (ABC) stacks
and, perhaps, even be used to discriminate the stacking sequence.\\
(vii) The AB-bilayer shows a remarkably small
value of the static dielectric constant of about $2.7$.
So small a value might be of interest for 
fast nano- or molecular electronics.\\
(viii) Although not investigated quantitatively in this paper,
it is obvious that in (ABC) slabs the strong dependence of the gap on the
perpendicular field will lead to
a strong dependence of the conductivity parallel to the layer
on the perpendicular field.\\
(ix) The local density of states was presented both for the
AB-bilayer and for an ABCAB-stack. 
This information can be useful for the identification
of a certain stack by infrared absorption spectroscopy.\\
(x) We found that
the screening properties of finite-thickness graphene stacks resemble
neither those of traditional metals nor those of traditional insulators.
For the AB-bilayer, the inter-layer charge transfer and
the induced atomic intra-layer dipole moments
contribute nearly equally to the total induced dipole moment.
In the thicker slabs, the screening charge per layer induced
by an external
field shows a rapid decay toward the interior of the slab.
More than $70\%$ of the induced charge is located 
in the very surface layers, where it is distributed roughly equally 
over the lattice sites within the layer cell.
On the other hand, the induced intra--layer dipole
moments decay on a much larger length-scale.
The latter contribution is not included in simplified 
screening models.
Consequently, such models can provide only qualitative
results. \\
(xi) Whereas (ABC)-stacked slabs show chemical charge transfer 
only near the surface, in the (AB) stacks this transfer
is largest in the interior and decreases toward the surface.
The electrons are transfered from the atoms with saturated $p_z$ bonds
toward atoms with dangling bonds.

As a bottom line, we note that (ABC) stacks rather than (AB) stacks
with more than two layers may turn out to be 
of potential interest for applications relying on the
tunability of electronic properties by an external electric field. 
Such slabs with rhombohedral stacking have non-metallic
bulk-like states up to a large critical thickness and
topologically protected zero-energy surface states.
Thus, they open a tunable gap in an external field which
can additionally be tailored by the choice of the layer thickness.
Beyond this practical aspect, many interesting features
are expected both for pure $\beta$-graphite micro-crystals
as well as for graphite with controlled embeddings of (ABC) stacks.

\section{Acknowledgements}
We gratefully acknowledge discussions with Helmut Eschrig,
with Jeroen van den Brink, and with Ulrike Nitzsche.
Financial support was provided by Deutsche Forschungsgemeinschaft
via grant RI932/6-1.

\end{document}